\documentstyle[emulateapj,epsfig]{article}
\received{}
\lefthead{PATTON ET AL.}
\righthead{DYNAMICALLY CLOSE GALAXY PAIRS IN SSRS2}

\begin{document}
\newcommand{\fb}{f_{b}}
\newcommand{\kmax}{k_{R_C\rightarrow B}^{\rm{max}}}
\newcommand{\vmax}{v^{\rm{max}}}
\newcommand{\rmax}{r^{\rm{max}}}
\newcommand{\rpmin}{r_p^{\rm{min}}}
\newcommand{\rpmax}{r_p^{\rm{max}}}
\newcommand{\rvmax}{r_v^{\rm{max}}}
\newcommand{\OII}{[O{\small\rm II}]}
\newcommand{\WOII}{$W_0$(O{\small\rm II})}
\newcommand{\OIIN}{[O{\small\rm II}]3727\AA}
\newcommand{\HD}{$H \delta$}
\newcommand{\WHD}{$W_0(H \delta )$}
\newcommand{\HDN}{$H\delta$4103\AA}
\def\text@#1{\mathchoice
  {\textdef@\displaystyle\f@size{#1}}%
  {\textdef@\textstyle\tf@size{\firstchoice@false #1}}%
  {\textdef@\textstyle\sf@size{\firstchoice@false #1}}%
  {\textdef@\textstyle \ssf@size{\firstchoice@false #1}}%
  \check@mathfonts
}
\def\textdef@#1#2#3{\hbox{{%
                    \everymath{#1}%
                    \let\f@size#2\selectfont
                    #3}}}
\def\cases{\left\{\def\arraystretch{1.2}\hskip-\arraycolsep
  \array{l@{\quad}l}}
\def\endcases{\endarray\hskip-\arraycolsep\right.}
\def\ion#1#2{#1$\;${\small\rm\@Roman{#2}}\relax}
\def\ion2#1{#1$\;${\small\rm\Roman{2}}\relax}
\def\arcmin{\hbox{$^\prime$}}
\def\arcsec{\hbox{$^{\prime\prime}$}}
\def\deg{\mbox{$^\circ$}}
\def\min{\mbox{$'$}}
\def\sec{\mbox{$''$}}
\def\secd{\mbox{$'' \!\! .$}}
\def\hrs{\mbox{$^{\rm h}$}}
\def\mins{\mbox{$^{\rm m}$}}
\def\secs{\mbox{$^{\rm s}$}}
\def\secsd{\mbox{$^{\rm s} \!\! .$}}
\def\lsim{\mbox{${\scriptstyle \buildrel < \over \sim}$}}
\def\like{{\cal L}}
\def\lesssim{\mathrel{\hbox{\rlap{\hbox{\lower4pt\hbox{$\sim$}}}\hbox{$<$}}}}
\def\gtrsim{\mathrel{\hbox{\rlap{\hbox{\lower4pt\hbox{$\sim$}}}\hbox{$>$}}}}
\let\la=\lesssim
\let\ga=\gtrsim
\newcommand{\frem}{f_{{\rm rem}}}
\newcommand{\fmg}{f_{{\rm mg}}}
\newcommand{\tmg}{T_{{\rm mg}}}
\newcommand{\rmg}{{\cal R}_{{\rm mg}}}
\newcommand{\rmgv}{{\cal R}_{{\rm mg}_{V}}}
\newcommand{\rmrv}{{\cal R}_{{\rm mr}_{V}}}
\newcommand{\rac}{{\cal R}_{{\rm ac}}}
\newcommand{\racv}{{\cal R}_{{\rm ac}_{V}}}
\newcommand{\arcsecpt}{\hbox to 1pt{}\rlap{\arcsec}.\hbox to 2pt{}}
\newcommand{\arcminpt}{\hbox to 1pt{}\rlap{\arcmin}.\hbox to 2pt{}}
\newcommand{\etal}{{\it et al.}~}
\newcommand{\Ic}{I$_{\rm{c}}$}
\newcommand{\hkpc}{$h^{-1}$~kpc}
\newcommand{\hmpc}{$h^{-1}$~Mpc}
\newcommand{\mbright}{M_{\rm{bright}}}
\newcommand{\mfaint}{M_{\rm{faint}}}
\newcommand{\mlim}{M_{\rm{lim}}}
\newcommand{\lsun}{L_{\odot}}
\newcommand{\msun}{M_{\odot}}
\newcommand{\eqsp}{\nonumber \\}
\submitted{Accepted for Publication in the Astrophysical Journal}
\title{
New Techniques for Relating Dynamically Close Galaxy Pairs to  
Merger and Accretion Rates : 
Application to the SSRS2 Redshift Survey}
\author{D. R. Patton\altaffilmark{1,2,3}, 
R. G. Carlberg\altaffilmark{3},
R. O. Marzke\altaffilmark{4}, 
C. J. Pritchet\altaffilmark{1}, 
L. N. da Costa\altaffilmark{5},
\& 
P. S. Pellegrini\altaffilmark{6}
}
\altaffiltext{1}{Department of Physics and Astronomy, 
University of Victoria,
PO Box 3055, Victoria, BC, V8W 3P6, Canada, patton, 
pritchet@uvastro.phys.uvic.ca}
\altaffiltext{2}{Guest User, Canadian Astronomy Data Centre, which 
is operated by the Herzberg Institute of Astrophysics, National 
Research Council of Canada.}
\altaffiltext{3}{Department of Astronomy, University of Toronto, 
60 St. George Street, Toronto, ON, 
M5S 3H8, Canada, patton, carlberg@astro.utoronto.ca}
\altaffiltext{4}{Observatories of the Carnegie Institute of 
Washington, 813 Santa Barbara St.,
Pasadena, CA 91101, USA, marzke@ociw.edu}
\altaffiltext{5}{European Southern Observatory,
Karl-Schwarzchild-Strasse 2, 85748 Garching bei Munchen, 
Germany, ldacosta@eso.org}
\altaffiltext{6}{Observat\'{o}rio Nacional, Rua General Jos\'{e} 
Cristino 77, 20921-030 S\~{a}o Crist\'{o}v\~{a}o, Rio de Janeiro,
Brazil, pssp@on.br}
\authoremail{patton@uvastro.phys.uvic.ca}

\begin{abstract}

The galaxy merger and accretion rates, and their evolution 
with time, provide 
important tests for models of galaxy formation and evolution. 
Close pairs of galaxies are the best available means of 
measuring redshift evolution in these quantities.
In this study, we introduce two new pair statistics, which 
relate close pairs to the merger and accretion rates.  
We demonstrate the importance of correcting these (and other) 
pair statistics for selection effects related to sample 
depth and completeness.  In particular, we highlight the 
severe bias that can result from the use of a flux-limited survey.  
The first statistic, denoted $N_c$, 
gives the number of companions per galaxy, within a specified 
range in absolute magnitude.
$N_c$ is directly related to the galaxy merger rate.
The second statistic, called $L_c$, 
gives the total luminosity in companions, per galaxy.  
This quantity can be used to investigate the mass accretion rate.  
Both $N_c$ and $L_c$ are 
related to the galaxy correlation function 
$\xi$ and luminosity function $\phi(M)$ in a straightforward 
manner.  Both statistics have been designed with selection 
effects in mind.  We outline techniques which account 
for various selection effects, and demonstrate the success 
of this approach using Monte Carlo simulations.  
If one assumes that clustering is independent of 
luminosity (which is appropriate for reasonable ranges in 
luminosity), then these statistics may be applied to flux-limited 
surveys.  

These techniques are applied to a sample of 5426 galaxies 
in the SSRS2 redshift survey.  This is the first 
large, well-defined low-$z$ survey to be used for pair statistics.  
Using close (5 \hkpc~ $\leq r_p \leq$ 20 \hkpc) 
dynamical ($\Delta v \leq 500$ km/s) pairs,
we find $N_c(-21 \leq M_B \leq -18)=0.0226 \pm 0.0052$ and 
$L_c(-21 \leq M_B \leq -18) = 0.0216 \pm 0.0055 \times 10^{10}~h^2 \lsun$ 
at $z$=0.015.  
These are the first secure estimates of low-redshift 
pair statistics, 
and they will provide local benchmarks for ongoing and future 
pair studies.
If $N_c$ remains fixed with redshift, 
simple assumptions imply that $\sim$ 6.6\% of present 
day galaxies with $-21 \leq M_B \leq -18$ have undergone 
mergers since $z$=1. 
When applied to redshift surveys of 
more distant galaxies, these techniques will yield 
the first robust estimates 
of evolution in the galaxy merger and accretion rates. 
\end{abstract}

\section{INTRODUCTION} \label{ssrs2mr:intro}

Studies of galaxy evolution have revealed surprisingly recent 
changes in galaxy populations.  Comparisons of present day 
galaxies with those at moderate ($z \sim 0.5$) and high 
($z \sim 3$) redshift have uncovered trends which are often 
dramatic, and may trace galaxies to the time at which they 
were first assembled into recognizable entities.   
These discoveries have shed new light on the formation of galaxies, 
and have provided clues as to the nature of their evolution.  
At $z < 1$, the picture that is emerging is one in which early type 
galaxies evolve slowly and passively, while late type galaxies 
become more numerous with increasing redshift (e.g., 
\cite{CNOC2LF}).  
At higher redshifts, deep surveys such as the Hubble 
Deep Field (\cite{HDF}) indicate an increase in the cosmic 
star formation rate out to $z \sim 2$ (e.g., 
Madau, Pozzetti, and Dickinson 1998\nocite{MADAU}).  

While considerable progress has been made in the observational 
description of galaxy evolution, important questions remain 
regarding the physical processes driving this evolution.  
Mechanisms that have been postulated include galaxy-galaxy 
mergers, luminosity-dependent luminosity 
evolution, and the existence of a new population of galaxies 
that has faded by the present epoch (see reviews by 
\cite{KK} and \cite{ELLIS97}).
In this study, we will investigate the relative importance of 
mergers in the evolution of field galaxies.  
Mergers transform the mass function of galaxies, marking a 
progression from small galaxies to larger ones.  
In addition, 
mergers can completely disrupt their constituent galaxies, 
changing gas-rich spiral galaxies into quiescent ellipticals  
(e.g., Toomre and Toomre 1972\nocite{TT72}).
During a collision, a merging system may also 
go through a dramatic transition, with the possible onset of 
triggered star formation and/or accretion onto a 
central black hole (see review by Barnes \& Hernquist 1992).   

It is clear that mergers do occur, even during the relatively 
quiet present epoch.  
However, the frequency of these 
events, and the distribution of masses involved, 
has yet to be accurately established.  This 
is true at both low and high redshift.  
Furthermore, while a number of attempts have been made, a secure 
measurement of evolution in the galaxy merger rate 
remains elusive, and a comparable measure of the accretion 
rate has yet to be attempted.    

In this study, we introduce a new approach for relating 
dynamically close galaxy pairs to merger and accretion 
rates.  These new techniques yield robust measurements
for disparate samples, thereby allowing meaningful 
comparisons of mergers at low and high redshift.  
In addition, these pair statistics 
can be adapted to a variety of redshift samples, and to 
studies of both major and minor mergers.  
We apply these techniques to a large sample of galaxies 
at low redshift (SSRS2), providing a much needed local benchmark 
for comparison with samples at higher redshift.    
In a forthcoming paper (Patton et al. 2000\nocite{P00}), we will apply 
these techniques to a large sample of galaxies at moderate 
redshift (CNOC2; $0.1 < z < 0.6$), 
yielding a secure estimate for the rate of evolution in the galaxy
merger and accretion rates.  

An overview of earlier pair studies, and a discussion of 
their limitations and shortcomings, 
are given in the next section.  
The SSRS2 data are described in \S~\ref{ssrs2mr:data}.  
Section~\ref{ssrs2mr:mrate} discusses the connection between
close pairs and the merger and accretion rates,
while \S~\ref{ssrs2mr:nclc} introduces new statistics for relating these 
quantities.  
Section~\ref{ssrs2mr:flux} describes how these statistics can be 
applied to flux-limited surveys in a robust manner.  
A pair classification experiment is presented in \S~\ref{ssrs2mr:class}, 
giving empirical justification for our close pair criteria.  
Pair statistics are then computed for the SSRS2 survey in 
\S~\ref{ssrs2mr:sample}, and the implications
are discussed in \S~\ref{ssrs2mr:discuss}.
Conclusions are given in the final section.  
Throughout this paper, we use a Hubble constant of 
$H_0 = 100 h$ km s$^{-1}$ Mpc$^{-1}$.  
We assume $h$=1 and $q_0$=0.1, unless stated otherwise.  

\section{BACKGROUND} \label{ssrs2mr:background}

Every estimate of evolution in the merger and/or accretion rate 
begins with the definition of a merger statistic.  Ideally, 
this statistic should be independent of selection effects 
such as 
optical contamination due to 
unrelated foreground/background galaxies, 
redshift incompleteness, 
redshift-dependent changes in minimum luminosity  
resulting from flux limits, 
contamination due to non-merging systems,
$k$-corrections, and luminosity evolution.  
In addition, 
it should be straightforward to relate the statistic to 
the global galaxy population, and to measurements on larger 
scales.  The statistic should then be applied to large, 
well-defined samples from low to high redshift, yielding 
secure estimates of how the merger and/or accretion rates 
vary with redshift.  

Within the past decade, there have been a number of attempts 
to estimate evolution in the galaxy merger rate using 
close pairs of galaxies
(e.g., \cite{ZK}, \cite{BKWF}, \cite{CPI}, 
\cite{WOODS}, \cite{YE}, \cite{P97}, \cite{LF99}).    
The statistic that has been most commonly employed is the 
traditional pair fraction, 
which gives the fraction of galaxies 
with suitably close physical companions.
This statistic is assumed to be proportional to the 
galaxy merger rate.  
The local (low-redshift) pair fraction was estimated by 
Patton et al. (1997), using a flux-limited ($B \leq 14.5$) 
sample of galaxies from the UGC catalog (\cite{UGC}).  
Using pairs with projected physical separations of less than 
20 \hkpc, they estimated the local pair 
fraction to be $4.3 \pm 0.4 \% $.
This result was shown to be consistent with the local pair 
fraction estimates of Carlberg et al. (1994) and Yee \& Ellingson 
(1995), both of whom also used the UGC catalog.  
The pair fraction has been measured for samples of galaxies 
at moderate redshift ($z \sim 0.5$), yielding published 
estimates ranging from approximately 0\% (\cite{WOODS}) 
to 34\% $\pm$ 9\% (\cite{BKWF}).  
Evolution in the galaxy merger rate is often parameterized 
as $(1+z)^m$.  Close pair studies have yielded a wide variety 
of results, spanning the range $0 \lesssim m \lesssim 5$.  
There are several reasons for the large spread in results.  
First, different methods have been used to relate the pair 
fraction to the merger rate.  In addition, some estimates have been 
found to suffer from biases due to 
optical contamination or redshift completeness.  
After taking all of these effects into account, Patton et al. (1997) 
demonstrated that most results are broadly consistent with their 
estimate of $m=2.8 \pm 0.9$, made using the largest redshift 
sample (545 galaxies) to date.  

While this convergence seems promising, 
all of these results have suffered from a number of 
very significant difficulties.  
The central (and most serious) problem has been the 
comparison between low and moderate redshift samples.  
Low-$z$ samples have been poorly defined, due to 
a lack of suitable redshift surveys.  
In addition, the pair fraction depends on 
both the clustering and mean density of galaxies.  The latter 
is very sensitive to the limiting absolute magnitude of 
galaxies, leading to severe redshift-dependent 
biases when using flux-limited galaxy samples.  
These biases have not been taken into account 
in the computation of pair fractions, or in the comparison between 
samples at different redshifts.  

While these problems are the most serious, there are several 
other areas of concern.  
A lack of redshift information has meant dealing with 
optical contamination due to unrelated foreground and background
galaxies.  Moreover, while one can statistically correct for 
this contamination, it is still not possible to discern low 
velocity companions from those that are physically associated 
but unbound, unless additional redshift information is available.   
Finally, there is no direct connection between  
the pair fraction and the galaxy correlation function (CF) 
and luminosity function (LF), making the results more 
difficult to 
interpret.

To address these issues, we have developed a novel 
approach to measuring pair statistics.  We will introduce 
new statistics that overcome many of the afflictions 
of the traditional pair fraction.  We will then apply these 
statistics to a large, well-defined sample of galaxies at 
low redshift.  

\section{DATA} \label{ssrs2mr:data}
The Second Southern Sky Redshift Survey 
(\cite{SSRS2}; hereafter SSRS2) consists of
5426 galaxies with $m_B \leq 15.5$, in two regions spanning a 
total of 1.69 steradians in the southern celestial hemisphere.  
The first region,
denoted SSRS2 South, has 
boundaries $-40 \deg \leq
\delta \leq -2.5 \deg$ and $b_{II} \leq -40 \deg$.  
The second region, SSRS2 North, is a
more recent addition, and is bounded by $\delta \leq 0 \deg$ and
$b_{II} \geq 35 \deg$.
Galaxies were selected primarily from the list of non-stellar 
objects in the {\it Hubble Space Telescope} Guide Star 
Catalog, with positions accurate to $\sim$ 1\arcsec~ and 
photometry with an rms scatter of $\sim$ 0.3 magnitudes (Alonso et
al. 1993\nocite{AL93}, Alonso et al. 1994\nocite{AL94}).  
Steps were taken to ensure that single 
galaxies were not mistakenly identified as close pairs, due to the 
presence of dust lanes, etc. (\cite{SSRS2}).  In addition, careful 
attention was paid to cases where a very close pair might be mistaken 
for a single galaxy.  This was found to make a negligible contribution
to the catalog as a whole ($< 0.1\%$ of galaxies are affected).  
The effect on the pairs analysis in this paper is further reduced 
by imposing a minimum pair separation 
of 5 \hkpc~(see \S~\ref{ssrs2mr:class}).  

The sample now includes redshifts for all galaxies brighter 
than $m_B \leq 15.5$.  
We correct all velocities to the local group barycenter 
using Equation 6 from Courteau and 
van den Bergh (1999)\nocite{VLG}.  
We restrict our analysis to the redshift range $0.005 \leq z \leq 
0.05$.  This eliminates nearby galaxies, for which recession 
velocities are dominated by peculiar 
velocities, giving poor distance estimates.   
We also avoid the sparsely sampled high redshift regime.
This leaves us with a well-defined sample of 4852 galaxies. 

\section{THE GALAXY MERGER RATE AND ACCRETION RATE} \label{ssrs2mr:mrate}
\subsection{Definitions} \label{ssrs2mr:mratedef}
The primary goal of earlier close pair studies has been to 
determine how the galaxy merger rate evolves with redshift.  
The merger rate affects the mass function of galaxies, 
and may also be connected to the cosmic star formation rate.  
Before attempting to measure the merger rate, it is important 
to begin with a clear definition of a merger and a merger rate. 
Here, we refer to mergers between two
galaxies which are both above some minimum mass or luminosity.  
If this minimum corresponds roughly to a typical bright galaxy
($L_*$), 
this criterion can be thought of as selecting so-called major 
mergers.   
We consider two merger rate definitions.  First, it is of interest 
to determine
the number of mergers that a typical galaxy will undergo per unit 
time.  In this case, the relevant rate may be termed the 
galaxy merger rate (hereafter $\rmg$).
A related quantity is the total number of mergers taking
place per unit time per unit co-moving volume.  We will refer 
to this as the volume merger rate (hereafter $\rmgv$).
Clearly, $\rmgv = n_0 \rmg$, where $n_0$ is the 
co-moving number density of galaxies.  

While both of these merger rates provide useful measures of 
galaxy interactions, they have their limitations.  
As one probes to faint luminosities, one will 
find an increasing number of faint companions; hence, 
the number of inferred mergers will increase in turn.
For all realistic LFs, this statistic will become 
dominated by dwarf galaxies.  
In addition, it is of interest to determine how 
the mass of galaxies will change due to mergers.  
To address these issues, we will also investigate the rate at 
which mass is being accreted onto a typical galaxy.  This quantity, 
the total mass accreted per galaxy per unit time, will 
be referred to as the galaxy accretion rate (hereafter $\rac$).  
This is related to the rate of mass accretion per unit co-moving 
volume ($\racv$) by $\racv = n_0 \rac$.
The mass (or luminosity) dependence of the accretion rate means 
that it will be dominated by relatively massive (or luminous) 
galaxies, with dwarfs playing a very minor role unless the 
mass function is very steep.  

\subsection{Observable Quantities} \label{ssrs2mr:observables}

In order to determine $\rmg$ observationally, 
one may begin by identifying systems which are destined to merge.  
By combining
information about the number of these systems and the timescale on
which they will undergo mergers, one can estimate an 
overall merger rate.  
Specifically, if one identifies $N_m$ ongoing mergers 
per galaxy, 
and if the average merging timescale for these systems 
is $\tmg$, then $\rmg = N_m/\tmg$.
If the mass involved in these mergers (per galaxy) is $M_m$, then 
$\rac = M_m/\tmg$.

In practice, direct measurement of these quantities is a daunting 
task.  It is difficult to determine if a 
given system will merge; furthermore, estimating the merger 
timescale for individual systems is challenging with the limited
information generally available.  However, if one simply wishes to
determine how the merger rate is {\it changing} with redshift, 
then the task is more manageable.
If one has the same definition of a merger in all samples under
consideration, then it is reasonable to assume that the 
merger timescale is the same for these samples.
In this case, we are left with the task of measuring 
quantities which are directly proportional to the 
number or mass of mergers per galaxy or per unit co-moving 
volume.  If one wishes to consider luminosity instead of mass, the 
relation between mass and luminosity must either be the same 
at all epochs, or understood well enough to correct for the 
differences.  

We have considered several quantities that fit this description.
All involve the identification of close physical associations 
of galaxies.    
A ``close companion'' is defined as a neighbour which will 
merge within a relatively short period of time 
($\tmg \ll t_H$), which allows an estimate of the 
instantaneous merger/accretion rate.  
If a galaxy is destined to undergo a merger in the very near 
future, it must have a companion close at hand.
One might attempt to estimate the number of mergers taking 
place within a sample of galaxies.  For example, a close pair 
of galaxies would be considered one merger, while a close triple 
would lead to two mergers, etc.  
Owing to the difficulty of determining with certainty which 
systems are undergoing mergers, we will not use this approach. 
One alternative is to estimate the number of galaxies
with one or more close companions, otherwise known as the 
pair fraction.  
One drawback of this approach is 
that close triples or higher order N-tuples complicate the 
analysis, since they are related to higher orders of the 
correlation function.  
This also makes it difficult to correct for the 
flux-limited nature of most redshift surveys.  
As a result, we choose to steer clear of 
this method also.  

In this study, we choose instead to use the {\it number and 
luminosity of close companions} per galaxy.  
The number of close companions per galaxy, hereafter $N_c$, 
is similar in nature to the pair fraction.
In fact, they are identical in a volume-limited sample with 
no triples or higher 
order N-tuples.  However, $N_c$ will prove to be much more 
robust and versatile.  
We assume that $N_c$ is directly proportional to 
the number of mergers per galaxy, such that $N_m = kN_c$ 
($k$ is a constant). 
This pairwise statistic is preferable to the number of 
mergers per galaxy or the fraction of galaxies in 
merging systems, in that 
it is related, in a direct and straightforward manner, to 
the galaxy two-point CF and the 
LF (see Section~\ref{ssrs2mr:nclc}).  
We note that it is not necessary that there 
be a one-to-one correspondence between companions and mergers, 
as long as the correspondence is the same, on average, 
in all samples under consideration.  

Using this approach to estimate the number of mergers per galaxy, 
the merger rate is then given by $\rmg = kN_c / \tmg$.   
The actual value of $k$ depends on the merging systems 
under consideration.  If 
one identifies a pure set of galaxy pairs, each definitely
undergoing a merger, then each pair, consisting of 2 companions, 
would lead to one merger, giving $k$=0.5.  For a pair 
sample which includes some triples and perhaps higher order 
N-tuples, $k < 0.5$.  If the merging sample under 
investigation contains some systems which are not truly merging 
(for instance, close pairs with hyperbolic orbits), then $k$ 
will also be reduced.  
While $k$ clearly varies with the type of merging system used, 
the key is for $k$ to be the same for all samples under 
consideration.

We take a similar approach with the accretion rate.  We again 
use close companions, and in this case we simply add up the 
luminosity in companions, per galaxy ($L_c$).  
Defining the mean companion 
mass-to-light ratio as $\Upsilon$, 
it follows that $M_m = \Upsilon L_c$ and 
$\rac = \Upsilon L_c / \tmg$.
When comparing different samples, 
any significant differences in $\Upsilon$ must be accounted for.  

\subsection{A Simple Model of Mass Function Evolution Due to Mergers}

In order to motivate further the need for merger rate measurements, 
and to set the stage for future work relating pair statistics to 
the mass and luminosity function, we develop a simple model 
which relates these important quantities.  Suppose the galaxy 
mass function is given by $\phi(M,t)$.  This function gives 
the number density of galaxies of mass $M$ 
at time $t$, per unit mass.  
The model that follows can also be expressed in terms 
of luminosity or absolute magnitude, rather than mass.  

We begin by assuming that all changes in the mass function 
are due to mergers.  While this is clearly simplistic, this 
model will serve to demonstrate the effects that various 
merger rates can have on the mass function.  In order to 
relate the mass function to the observable luminosity function, 
we further assume that mergers do not induce star formation.  
Again, this is clearly an oversimplification; however, this 
simple case will still provide a useful lower limit on the 
relative contribution of mergers to LF evolution.  
Finally, we assume that merging is a binary process.  

Following \cite{BT88}, we consider how $\phi(M,t)$ 
evolves as the universe ages 
from time $t$ to time $t+\Delta t$.  Each merger will remove  
two galaxies from the mass function, and produce one new galaxy. 
Let $\Delta \phi(M,\Delta t)_{{\rm out}}$ represent the 
decrease in the mass function due to galaxies removed by mergers, 
while $\Delta \phi(M,\Delta t)_{{\rm in}}$ gives the increase 
due to the remnants produced by these mergers.  
Evolution in the mass function can then be given by 
\begin{equation}\label{ssrs2mr:eqinout}
\phi(M,t+\Delta t) = \phi(M,t) 
- \Delta \phi(M,\Delta t)_{{\rm out}}
+ \Delta \phi(M,\Delta t)_{{\rm in}}. 
\end{equation}

We model this function by considering all galaxy pairs, 
along with an expression for the merging likelihood of each.  
Let $p(M,M^{\prime},\Delta t)$ denote the probability 
that a galaxy of mass $M$ will merge with a galaxy of mass 
$M^{\prime}$ in time interval $\Delta t$. 
In order to estimate $\Delta \phi(M,\Delta t)_{{\rm out}}$, 
we need to take all galaxies of mass $M$, and integrate over 
all companions, yielding
\begin{eqnarray}\label{ssrs2mr:eqout}
\Delta \phi(M,\Delta t)_{{\rm out}} = \eqsp
\int_0^{\infty} 
\phi(M,t) \phi(M^{\prime},t) p(M,M^{\prime},\Delta t)
[h^{-1}{\rm Mpc}]^3dM^{\prime}.
\end{eqnarray}
We devise a comparable expression for 
$\Delta \phi(M,\Delta t)_{{\rm in}}$ by integrating over 
all pairs with end$-$products of mass $M$.  This is achieved 
by considering all pairs with component of 
mass $M-M^{\prime}$ and $M^{\prime}$, such that
\begin{eqnarray}\label{ssrs2mr:eqin}
\Delta \phi(M,\Delta t)_{{\rm in}} = 
\int_0^{\infty}
\phi(M-M^{\prime},t) \phi(M^{\prime},t) \eqsp
\times p(M-M^{\prime},M^{\prime},\Delta t)
[h^{-1}{\rm Mpc}]^3dM^{\prime}.
\end{eqnarray}

We can also express Equation~\ref{ssrs2mr:eqinout} in terms of 
the pair statistics outlined in Section~\ref{ssrs2mr:observables}. 
If one considers close companions of mass $M^{\prime} \leq \infty$ 
next to primary galaxies of mass $M$, the volume merger rate 
can be expressed as $\rmgv (M)$, yielding 
\begin{equation}
\Delta \phi(M,\Delta t)_{{\rm out}} = \rmgv (M)\Delta t.
\end{equation}
Similarly, if one defines a merger remnant statistic, $\rmrv (M)$,
to be the co-moving number density of merger remnants 
per unit time corresponding to these same mergers, then 
\begin{equation}
\Delta \phi(M,\Delta t)_{{\rm in}} = \rmrv (M)\Delta t.
\end{equation}
Therefore, it is possible, in principle, 
to use pair statistics to measure the evolution in the 
mass or luminosity function due to mergers.  
However, current pair samples are too small to permit useful 
pair statistics for different mass combinations.  In addition, 
present day observations of close pairs are not of sufficient 
detail to determine the proportion of pairs that will result 
in mergers (factor $k$ in previous section).  
Moreover, timescale estimates for these mergers are not known with 
any degree of certainty.  
Hence, useful observations of mass function evolution due to 
mergers will have to wait for improved pair samples and 
detailed estimates of merger timescales.  

\section{A NEW APPROACH TO MEASURING PAIR STATISTICS} \label{ssrs2mr:nclc}
In this section, we outline the procedure for measuring 
the mean number ($N_c$) and luminosity ($L_c$) of close companions 
for a sample of galaxies with measured redshifts.  
We begin by 
defining these statistics in real space, demonstrating how 
they are related to the galaxy LF and CF.  
We then show how these statistics can be applied 
in redshift space.  

\subsection{Pair Statistics in Real Space}\label{ssrs2mr:realspace}
In this study, we will measure pair statistics for a complete 
low-redshift sample of galaxies (SSRS2).  
However, we wish to make these statistics applicable to a 
wide variety of redshift samples.  
We would also like this method to be useful for 
studies of minor mergers, where one is interested in faint 
companions around bright galaxies.   
Moreover, these techniques should be adaptable
to redshift samples with varying degrees of completeness (that is, 
with redshifts not necessarily available for every galaxy).   
Therefore, in the following analysis, we treat host galaxies 
and companions differently.  

Consider a primary sample of $N_1$ host galaxies with absolute 
magnitudes $M \leq M_1$, lying in some volume $V$.  Suppose 
this volume also contains a secondary sample of $N_2$ galaxies 
with $M \leq M_2$.  In the general case, the primary and 
secondary samples may have galaxies in common.  This includes 
the special case in which the two samples are identical.  
If $M_1 \approx M_2$, this will tend to probe major mergers.  
If $M_2$ is chosen to be significantly fainter than $M_1$, 
this will allow for the study of minor mergers.  
We assume here that both samples are complete to the given 
absolute magnitude limits; in Section~\ref{ssrs2mr:flux}, we extend 
the analysis from volume-limited samples to those that are 
flux-limited.

We wish to determine the mean number and luminosity of 
companions (in the secondary sample) for galaxies in the 
primary sample.  In real space, we define a close companion 
to be one that lies at a true physical separation 
of $r \leq \rmax$, where $\rmax$ is some appropriate 
maximum physical separation.  
To compute the observed mean number ($N_c$) and luminosity ($L_c$) 
of companions, we simply add up the number ($N_{c_i}$) and 
luminosity ($L_{c_i}$) of companions for each of the $N_1$ 
galaxies in the primary sample, and then compute the mean.  
Therefore, 
\begin{equation}\label{ssrs2mr:eqncbasic}
N_c (M \leq M_2) = \sum_{i=1}^{N_1} N_{c_i}/N_1
\end{equation}
and 
\begin{equation}\label{ssrs2mr:eqlcbasic}
L_c (M \leq M_2) = \sum_{i=1}^{N_1} L_{c_i}/N_1.
\end{equation}

We can also estimate what these statistics should be, given 
detailed knowledge of the 
galaxy two-point CF $\xi$ and the 
LF $\phi(M)$. 
This is necessary if one wishes to relate these pair statistics to 
measurements on larger scales.
Consider a galaxy in the primary sample with absolute magnitude 
$M_i$ at redshift $z_i$.  We would first like to estimate the 
number and luminosity of companions lying in a shell 
at physical distance ({\it proper} co-ordinates)  
$r \leq r_{ij} \leq r+dr$ from this primary galaxy.  
To make this estimate, we need to know the mean density 
of galaxies (related to the LF), and the 
expected overdensity in the volume of interest (given by 
the CF).  
The mean physical number density of galaxies at redshift $z_i$ 
in the secondary sample, with absolute magnitudes 
$M \leq M_j \leq M + dM$, is given by 
\begin{equation}\label{ssrs2mr:eqn2}
n_2(z_i,M)dM = (1+z_i)^3 \phi(M,z_i)dM, 
\end{equation}
where $\phi(M,z_i)$ 
is the differential galaxy LF, which specifies the {\it co-moving} 
number density of galaxies at redshift $z_i$, in units 
of $h^3{\rm Mpc}^{-3}{\rm mag}^{-1}$.
The actual density of objects in the region of interest  
is determined by multiplying the mean density by (1+$\xi$), 
where $\xi$ is the overdensity given by the two-point 
CF (\cite{PJE1}).  
In general, $\xi$ depends on the pair separation $r$, 
the mean redshift $z_i$, 
the absolute magnitude of each galaxy ($M_i$, $M_j$), and the 
orbits involved, specified by components parallel ($v_{\|}$) and 
perpendicular ($v_{\bot}$) to the line of sight.   
It follows that the mean number of companions 
with $M \leq M_j \leq M + dM$ and 
$r \leq r_{ij} \leq r+dr$ 
is given by 
\begin{eqnarray} \label{ssrs2mr:ncdiff}
N_{c_i}(z_i,M_i,M,r,v_{\bot},v_{\|})dMdr = \eqsp
n_2(z_i,M)dM [1+\xi(r,z_i,M_i,M,v_{\bot},v_{\|})]4\pi r^2dr.
\end{eqnarray}

We must now integrate this expression for all companions 
with $M \leq M_2$ and $r \leq \rmax$.  
Integration over the LF yields 
\begin{equation}\label{ssrs2mr:eqn2int}
n_2(z_i,M \leq M_2) = (1+z_i)^3 \int_{-\infty}^{M_2}\phi(M,z_i)dM. 
\end{equation}
Integration over the CF is non-trivial, because of 
the complex nature 
of $\xi(r,z_i,M_i,M,v_{\bot},v_{\|})$.  
With redshift samples that are currently available, 
it is not possible to measure this 
dependence accurately for the systems of interest.  
Hence, we must make three important assumptions at this stage.  
First, we assume that $\xi$ is independent of luminosity.  
Later in the paper, we demonstrate empirically that this is 
a reasonable assumption, provided one selects a sample with 
appropriate ranges in absolute magnitude 
(see Section~\ref{ssrs2mr:clust}).  
Secondly, we assume that the distribution of velocities is 
isotropic.  If one averages over a reasonable number of pairs, 
this is bound to be true, and therefore $\xi$ is independent 
of $v_{\bot}$ and $v_{\|}$.  
Finally, we assume that the form of the CF, as 
measured on large scales, can be extrapolated to the small scales 
of interest here.  
This assumption applies only to the method 
of relating pairs to large scale measures, and not to the 
actual measurement of pair statistics.

It is now straightforward to integrate Equation~\ref{ssrs2mr:ncdiff}.  
The mean number of companions with $M \leq M_2$ and 
$r \leq \rmax$ for a primary galaxy at redshift $z_i$ is given by 
\begin{eqnarray} \label{ssrs2mr:ncint}
N_{c_i}(z_i,M\leq M_2,r\leq \rmax) = \eqsp
n_2(z_i,M \leq M_2)
\int_0^{\rmax}[1+\xi(r,z_i)]4\pi r^2dr.
\end{eqnarray}
We derive an analogous expression for the mean
luminosity in companions.  The integrated luminosity density is 
given by 
\begin{equation}\label{ssrs2mr:eql2int}
j_2(z_i,M \leq M_2) = (1+z_i)^3 
\int_{-\infty}^{M_2}\phi(M,z_i)L(M)dM, 
\end{equation}
where 
\begin{equation}\label{ssrs2mr:eqL}
L(M) = 10^{-0.4(M-\msun)}\lsun.
\end{equation}
Therefore, 
\begin{eqnarray} \label{ssrs2mr:lcint}
L_{c_i}(z_i,M\leq M_2,r\leq \rmax) = \eqsp
j_2(z_i,M \leq M_2)
\int_0^{\rmax}[1+\xi(r,z_i)]4\pi r^2dr.
\end{eqnarray}
Given measurements of the CF on large scales, it is then 
straightforward to integrate these equations to arrive 
at predicted values of $N_{c_i}$ and $L_{c_i}$.  

It is important to note that these statistics are directly 
dependent on $M_2$, which affects the mean density of galaxies 
in the secondary sample.  
This is different from statistics such as 
the correlation function, which are independent of density.  
Hence, this serves as a reminder that we must exercise caution
when choosing our samples, to ensure that differences in 
the pair statistics (and hence in the merger and 
accretion rates) are not 
simply due to apparent density differences resulting from 
selection effects.  In addition, note that the choice of 
$M_1$ has no density-related effects on $N_c$ and $L_c$.  

\subsection{Dynamical Pairs in Redshift Space}\label{ssrs2mr:zspace}
While it is preferable to identify companions 
based on their true physical pair separation, this 
is clearly not feasible when dealing with data from 
redshift surveys.  In the absence of independent distance 
measurements for each galaxy, one must resort to identifying 
companions in redshift space.  In this section, we outline 
a straightforward approach for measuring our new pair statistics 
in redshift space.  We then attempt to relate these statistics 
to their counterparts in real space.

For any given pair of galaxies in redshift space, one can 
compute two basic properties which describe the intrinsic 
pair separation : the projected physical 
separation (hereafter $r_p$) and the rest-frame 
relative velocity along the line of sight (hereafter $\Delta v$).   
For a pair of galaxies with redshifts 
$z_i$ (primary galaxy) and $z_j$ (secondary), with angular 
separation $\theta$, these quantities are given by 
$r_p = \theta d_A(z_i)$
and $\Delta v = c |z_j~-~z_i| / (1+z_i)$, where 
$d_A(z_i)$ is the angular diameter distance at redshift $z_i$.  
Note that $r_p$ gives the projected separation at the 
redshift of the primary galaxy.  

We define a close companion as one in which 
the separation (both projected and line-of-sight) 
is less than some appropriate separation, such that 
$r_p \leq \rpmax$ and $\Delta v \leq \Delta \vmax$. 
The line-of-sight criterion depends on both the 
physical line-of-sight separation and the line-of-sight 
peculiar velocity of the companion.  It is of course not 
possible to determine the relative contributions of these 
components without distance information.  However, for 
the small companion separations we will be concerned with, 
the peculiar velocity component is likely to be dominant 
in most cases, as we will be dealing with a field sample 
of galaxies (the same would not be true in the high velocity 
environment of rich clusters).  Hence, this criterion serves primarily to 
identify companions with low peculiar velocities.  While 
this is fundamentally different from the pure separation 
criterion used in real space, it too will serve to identify 
companions with the highest likelihood of undergoing 
imminent mergers.
Using this definition of a close companion, it is straightforward 
to compute $N_c$ and $L_c$, using Equations~\ref{ssrs2mr:eqncbasic} 
and \ref{ssrs2mr:eqlcbasic}. 
Thus, the complexities of redshift space do not greatly 
complicate the computation of these pair statistics.  

As in real space, we wish to relate these statistics 
to measurements on larger scales, 
given reasonable assumptions about the LF and CF.  
The situation is more complicated in redshift space, and 
therefore involves additional assumptions.  We stress, however, 
that these assumptions apply only to the method of relating 
pair statistics to large scale measures, and
not to the measured pair statistics themselves.  
To outline an algorithm for generating these predictions, 
we follow the approach of the previous section.  
We begin by modifying Equations~\ref{ssrs2mr:ncint} and ~\ref{ssrs2mr:lcint}, 
integrating over the new pair volume defined in redshift space.  
In order to do this, we use the two dimensional correlation 
function in redshift space, $\xi_2(r_p,r_v,z)$, giving 
\begin{eqnarray} \label{ssrs2mr:nc2int}
N_{c_i}(z_i,M\leq M_2,r_p \leq \rpmax,| r_v | \leq \rvmax) = \eqsp
n_2(z_i,M \leq M_2)
\int_{-\rvmax}^{\rvmax}\int_0^{\rpmax}[1+\xi_2(r_p,r_v,z_i)]2\pi r_pdr_pdr_v
\end{eqnarray}
and 
\begin{eqnarray} \label{ssrs2mr:lc2int}
L_{c_i}(z_i,M\leq M_2,r_p \leq \rpmax,| r_v | \leq \rvmax) = \eqsp
j_2(z_i,M \leq M_2)
\int_{-\rvmax}^{\rvmax}\int_0^{\rpmax}[1+\xi_2(r_p,r_v,z_i)]2\pi r_pdr_pdr_v.
\end{eqnarray}
  
The two dimensional correlation function 
is the convolution of the velocity distribution 
in the redshift direction, $F(v_z)$, with the spatial correlation 
function $\xi(r,z)$, given by 
\begin{equation}\label{ssrs2mr:eqxi2}
\xi_2(x_p,x_v,z) = \int_{-\infty}^{\infty}\xi(\sqrt{x_p^2+y^2},z)
F(H(z)[y-x_v])dy.
\end{equation}
Here, $H(z)$ is the Hubble constant at redshift $z$, 
given by $H(z) = H_0(1+z)\sqrt{1+\Omega_0 z}$.  
We have ignored the effect of
infall velocities, which must be taken into account at larger 
radii but is an acceptable approximation for small separations. 
If the form of the CF and LF are known, it is straightforward 
to integrate Equations~\ref{ssrs2mr:nc2int} and \ref{ssrs2mr:lc2int}, yielding 
predictions of $N_c$ and $L_c$.  

\subsection{Physical Pairs in Redshift Space}\label{ssrs2mr:pspace}

It is not always possible to have precise redshifts for all 
galaxies of interest in a sample.  A common scenario with redshift 
surveys is to have redshifts available for a subset of galaxies 
identified in a flux-limited photometric sample.  
The photometric sample used to select galaxies for follow-up 
spectroscopy probes to fainter apparent magnitudes than the 
spectroscopic sample.  
In addition, the spectroscopic sample may be incomplete, even 
at the bright end of the sample.  
In this section, we will describe the procedure for applying 
pair statistics to this class of samples.  

Suppose the primary sample is defined as all galaxies in the 
spectroscopic sample with absolute magnitudes $M \leq M_1$.
The secondary sample consists of all galaxies lying in the 
photometric sample, regardless of whether or not they have 
measured redshifts.  
Once again, there may be some 
overlap between the primary and secondary samples.  
We must now identify close pairs.  For each primary-secondary 
pair, we can compute $r_p$ in precisely the same manner as 
before (see previous section), since we need only the redshift 
of the primary galaxy and the angular separation of the pair.  
However, we are no longer able to compute the relative 
velocity along the line of sight, since this requires redshifts 
for both members of the pair.  Thus, we do not have enough 
information to identify close dynamical pairs.  However, 
it is still possible to determine, in a statistical manner, 
how many physically associated companions are present.  This 
is done by comparing the number (or luminosity) of 
observed companions with the number (or luminosity) 
expected in a random distribution.  

As stressed in Section~\ref{ssrs2mr:realspace}, pair statistics depend 
on the minimum luminosity $M_2$ imposed on the secondary sample.  
While we are now unable to compute the actual luminosity for 
galaxies in the secondary sample, we must still impose $M_2$ 
if the ensuing pair statistics are to be meaningful.  
To do this, we make use of the fact that all physical companions 
must lie at approximately the same redshift as the primary galaxy 
under consideration.  Therefore, $M_2$ corresponds to a limiting 
{\it apparent} magnitude $m_2$ at redshift $z_i$, such that 
\begin{equation}\label{ssrs2mr:eqm2}
m_2  = M_2(z_i) - 5 \log h + 25 + 5 \log d_L(z_i) + k(z_i),
\end{equation}
where $d_L(z_i)$ is the luminosity distance at redshift $z_i$, 
and $k(z_i)$ is the $k$-correction.  

To begin, one finds all observed close companions with 
$m \leq m_2(z_i)$, using 
only the $r_p$ criterion.  This results in the quantities 
$N_c^{\rm D}$ and $L_c^{\rm D}$, where the ``D'' superscript 
denote companions found in the data sample.  
One must then estimate the number ($N_c^{\rm R}$) and 
luminosity ($L_c^{\rm R}$) of companions 
expected at random.  
The final pair statistics for close physical companions are 
then given by 
$N_c = N_c^{\rm D} - N_c^{\rm R}$ and 
$L_c = L_c^{\rm D} - L_c^{\rm R}$.

We will now describe how to predict these statistics using the 
known LF and CF.  This is relatively straightforward, since 
the excess $\xi$ given by the CF is determined by the relative 
proportions of real and random companions.  
The pair statistics are once again integrals over the 
two dimensional CF in redshift space, as specified by
Equations~\ref{ssrs2mr:nc2int} and \ref{ssrs2mr:lc2int}.
In the ``$1+\xi$'' term, the first part gives the random 
contribution, while the second gives the excess over random.  
Thus, these pair statistics give the true density of companions, 
rather than the ``excess'' density.  This is intentional, 
since mergers will occur even in an uncorrelated, randomly 
distributed sample of galaxies.  At the small separations of 
interest, usually less than 1\% of the correlation length, 
the difference between the mean density and the mean overdensity 
is less than about 0.01\% in real space.  For practical 
measurements in redshift space, where $r_v = \Delta v H(z)^{-1}$ 
is of order the correlation length, the background contribution 
is substantially larger than real space, but still amounts to 
less than 1\%.  
Thus, for the close pairs considered in this study, 
it is reasonable to ignore the contribution that random 
companions make to the sample of physical companions.  
That is, we take $1+\xi \approx \xi$.
This allows us to relate the predictions to the measured 
pair statistics set out above.  
In principle, Equations~\ref{ssrs2mr:nc2int} and 
\ref{ssrs2mr:lc2int} can be integrated 
over the range $-\infty < r_v < \infty$ to 
obtain predictions of $N_c$ and $L_c$. 
  
\subsection{Application to Volume-limited Monte Carlo Simulations} 
\label{ssrs2mr:MC}
To illustrate the concepts introduced so far, 
and to emphasize how these statistics depend on $M_2$, 
we apply these techniques to volume-limited 
Monte Carlo simulations, 
which mimic the global distribution of galaxies in the 
SSRS2 North and South catalogs.  
Using $q_0$=0.5, galaxies are distributed randomly 
within the co-moving volume enclosed by 
$0.005 \leq z \leq 0.05$ and the SSRS2 boundaries on the 
sky (see Section~\ref{ssrs2mr:data}). 
All peculiar velocities are set to zero.
To create a volume-limited sample, we impose a minimum 
luminosity of $M_B$=$-16$, and 
assign luminosities using the SSRS2 LF (\cite{ROM98}), 
which has 
Schechter function parameters 
$M_* - 5 \log h$=$-19.43$, 
$\alpha$=$-1.12$, and 
$\phi_*$ = $0.0128~ h^3 {\rm Mpc}^{-3} {\rm mag}^{-1}$.   
An arbitrarily large number of galaxies can be generated, 
which is of great assistance when looking for small 
systematic effects.
We produce 16000 galaxies in the South, and 8070 in the North;  
this gives the same density of galaxies in both regions.  

Using these simulations, 
we compute $N_c$ and $L_c$.  
As these galaxies are distributed randomly (as 
opposed to real galaxies which are clustered), close pairs 
are relatively rare.  
To ensure a reasonable yield of pairs, we use a pair 
definition of $\rpmax$=1 \hmpc~ and 
$\Delta \vmax$=1000 km/s.  We note that there are no peculiar 
velocities in these simulations; hence, the $\Delta \vmax$ 
criterion provides upper and lower limits on the line-of-sight 
distance to companions.
Also, recall from the preceding section that the choice of 
$M_1$ has no effect on the pair statistics if clustering is 
independent of luminosity.  
Hence, we choose $M_1$=$M_2$, which maximizes the 
size of the primary sample, and therefore minimizes the 
measurement errors in $N_c$ and $L_c$.  

\begin{figure}
\unitlength1.0cm
\begin{center}
\begin{picture}(16.0,16.0)
\put(0,0)
{\epsfig{file=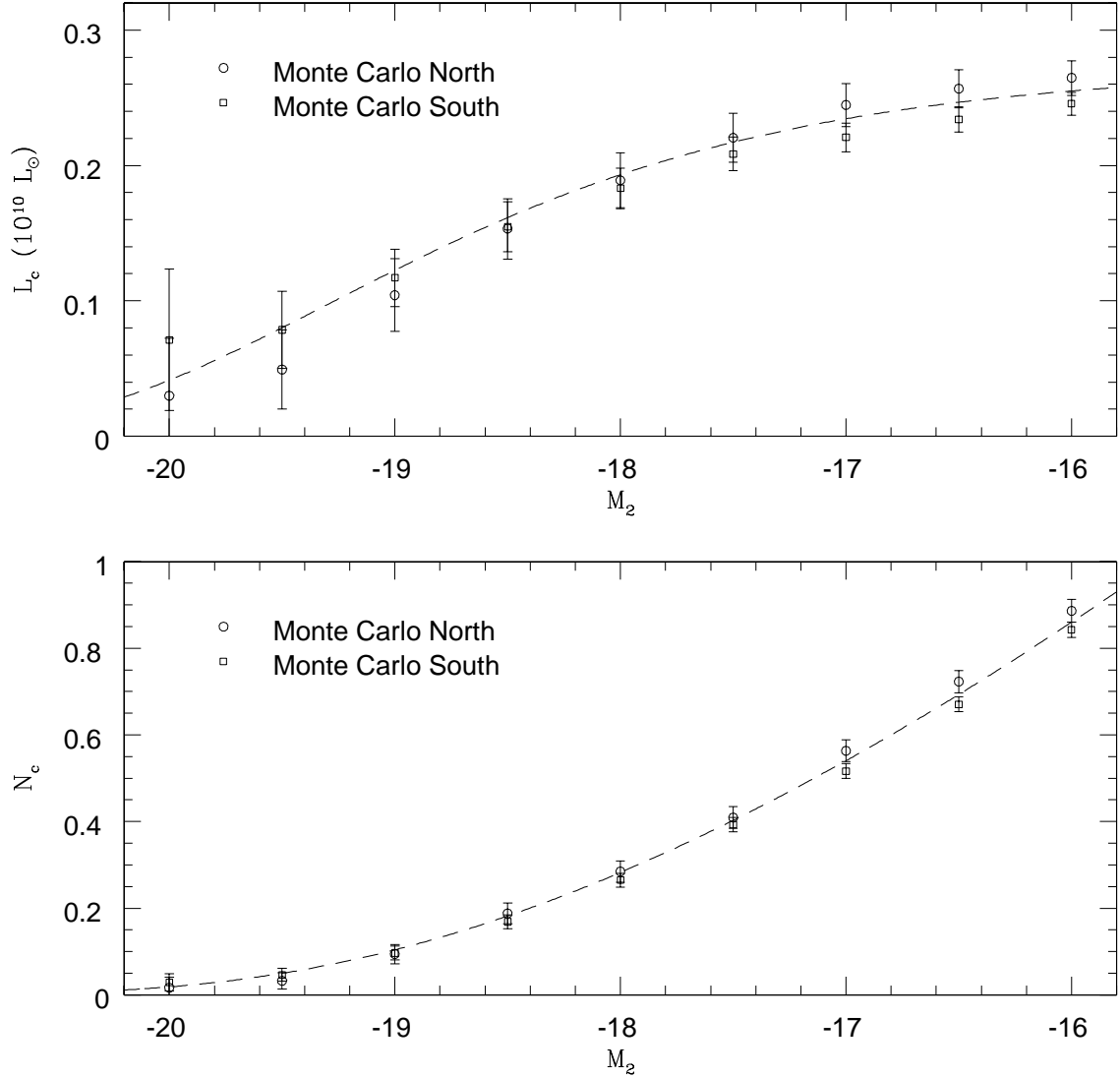,height=16.0cm,width=16.0cm}}
\end{picture}
\end{center}
\figcaption[figm2vl.ps]
{
Pair statistics are computed for the volume-limited Monte Carlo 
simulations.  $N_c$ and $L_c$ are given for a range of choices 
of $M_2$, with $M_1$=$M_2$.  
Error bars are computed using the Jackknife technique.
The dashed lines are the relations 
predicted using the input SSRS2 luminosity function, and 
normalized to match the data at $M_2$=$-16$.  
\label{ssrs2mr:figm2vl}}
\end{figure}

With these assumptions, we compute pair statistics
for a range of choices of $M_2$.  
Errors are computed using the Jackknife technique.  
For this resampling method, partial standard deviations, 
$\delta_i$, are computed for each object by taking the 
difference between the quantity being measuring, $f$, 
and the same quantity with the $i^{\rm{th}}$ galaxy removed 
from the sample, $f_i$, such that $\delta_i = f-f_i$.  
For a sample of $N$ galaxies, the variance is given by 
$[(N-1)/N\sum_i\delta_i^2]^{1/2}$ 
(Efron 1981\nocite{JACK1}; 
Efron \& Tibshirani 1986\nocite{JACK2}).  
Results are given in Figure~\ref{ssrs2mr:figm2vl}.  
Both statistics continue to increase as $M_2$ becomes 
fainter.  
$N_c$ diverges at faint magnitudes, while $L_c$ is seen to 
converge.  
This behaviour is a direct consequence of the 
shape of the LF; $N_c$ converges for $\alpha > -1$, while 
$L_c$ converges for $\alpha > -2$.  
The existence and magnitude of these trends  
clearly demonstrate the need to specify $M_2$ when 
measuring pair statistics.  

\section{APPLICATION TO FLUX-LIMITED SAMPLES}\label{ssrs2mr:flux}
The preceding section gives a straightforward prescription for computing 
pair statistics in volume-limited samples.  
However, redshift surveys are generally flux-limited.
By defining a volume-limited sample within such a survey, 
one must discard a large proportion of the data.  
In this section, we will outline how these pair statistics can 
be applied to flux-limited surveys.  

Pair statistics necessarily depend on both clustering and 
mean density, as 
shown by Equations~\ref{ssrs2mr:nc2int} and \ref{ssrs2mr:lc2int}.
In a flux-limited sample, both clustering and mean density will 
vary throughout the sample.  
We will use these equations to account for 
redshift-dependent changes in mean density,  
and we will demonstrate how to minimize the effects of 
clustering differences.  These techniques will then be 
tested with Monte Carlo simulations.  

\subsection{Dependence on Clustering} \label{ssrs2mr:clust}
By removing the fixed luminosity limit, 
the overall distribution of galaxy luminosities will vary 
with redshift within the sample, and the mean luminosity of 
the sample will differ from the volume-limited sample.   
However, galaxy clustering is known to be luminosity dependent.   
Measures of the galaxy correlation function (e.g., Loveday et al. 
1995, Willmer et al. 1998\nocite{W98}), 
power spectrum (e.g., Vogeley 1993\nocite{V93}), and counts in cells 
(Benoist et al. 1996) all find that luminous galaxies ($L > L_*$)
are more clustered than sub-$L_*$ galaxies, typically by 
a factor of $\sim 2$.  
This increase in clustering may be particularly strong 
(factor $\sim 4$) for very luminous galaxies ($M_B < -21$).

Clearly, this effect should not be ignored when computing pair 
statistics.  In principle, this could be incorporated into the 
measurement of these pair statistics.  However, available pair 
samples are too small to measure this dependence.  
We choose instead to minimize these effects 
by restricting the analysis to a fixed range in absolute 
magnitude, within which luminosity-dependent clustering is 
small or negligible.  
This is done by imposing additional bright ($\mbright$) 
and faint ($\mfaint$) 
absolute magnitude limits on the sample.  
Having thereby reduced the effects of luminosity segregation, we then assume
that the remaining differences will not have a significant 
effect on the measured pair statistics.  In Section~\ref{ssrs2mr:sense}, 
we use the SSRS2 sample to demonstrate empirically that this is in 
fact a reasonable assumption.  

\subsection{Dependence on Limiting Absolute Magnitude}
In Section~\ref{ssrs2mr:nclc}, we demonstrated that these pair 
statistics are meaningful only if one specifies the minimum 
luminosity of the primary and secondary samples.  For a 
flux-limited sample, however, the minimum luminosity of 
the sample increases with redshift.  One must therefore 
decide on a representative minimum luminosity, and account 
for differences between the desired minimum luminosity and 
the redshift-dependent minimum imposed by the apparent 
magnitude limit of the sample.  
If the LF is known, this can be achieved by 
weighting each galaxy appropriately.  
In this section, we outline a weighting scheme 
which makes this correction.  

\subsubsection{Weighting of Secondary Sample} \label{ssrs2mr:w2}
Consider a flux-limited sample in which host galaxies are 
located at a variety of redshifts.  Those at low redshift 
will have the greatest probability of having close companions 
that lie above the flux limit, since the flux limit corresponds 
to an intrinsic luminosity that is fainter than that for galaxies 
at higher redshift.  If we wish to avoid an inherent bias 
in the pair statistics, we must correct for this effect.  
Furthermore, we must account for any limits in absolute magnitude 
imposed on the sample to reduce the effects of luminosity-dependent 
clustering (\S~\ref{ssrs2mr:clust}).  
Finally, we have demonstrated the importance of specifying 
a limiting absolute magnitude for companions ($M_2$) when 
computing pair statistics.  Therefore, we must attempt to 
correct the pair statistics to the values that would have been 
achieved for a volume-limited secondary sample 
with $\mbright \leq M \leq M_2$.
Qualitatively, this correction 
should assign greater importance (or weight) to the more rare 
companions found at the high redshift end of the flux-limited 
sample.  
To make this correction as rigorous as possible, we will use 
the galaxy LF.  By integrating the LF over a given range in  
absolute magnitude, one can obtain an estimate of the mean 
number or luminosity density of galaxies in the sample.  
By performing this integration at any given redshift, 
accounting for the allowed ranges in absolute magnitude and 
the flux limit, it is possible to quantify
how the mean density varies with redshift within the 
defined sample.  This 
information can be used to remove this unwanted bias from 
the pair statistics.  

We assign a weight to each galaxy in the secondary sample, which 
renormalizes the sample 
to the density corresponding to $\mbright \leq M \leq M_2$.  
We must first determine $\mlim(z_i)$, which gives the 
limiting absolute magnitude allowed at redshift $z_i$.  At most 
redshifts, this is imposed by the limiting apparent 
magnitude $m$, 
such that $\mlim (z_i) = m -5\log d_L(z) - 25 - k(z)$.  
At the low redshift end of the sample, however, 
$\mfaint$ (defined in \S~\ref{ssrs2mr:clust}) will take over.  
That is, 
the limiting absolute magnitude used for identifying galaxies 
in the secondary sample is given by
\begin{equation}\label{ssrs2mr:eqmlim}
\mlim (z_i) = \max [M_{\rm{faint}},m -5\log d_L(z) - 25 - k(z)].
\end{equation}
The selection function, denoted $S(z)$, is defined as the 
ratio of densities in flux-limited versus volume-limited 
samples.  This function, given in terms of number density 
($S_N(z)$) and luminosity density ($S_L(z)$), is as 
follows : 
\begin{equation} \label{ssrs2mr:eqwN2}
S_{N}(z_i) = {\int_{\mbright}^{\mlim(z_i)} \phi(M)dM \over 
\int_{\mbright}^{M_2} \phi(M)dM}
\end{equation}
\begin{equation} \label{ssrs2mr:eqwL2}
S_{L}(z_i) = {\int_{\mbright}^{\mlim(z_i)} \phi(M)L(M)dM \over 
\int_{\mbright}^{M_2} \phi(M)L(M)dM},
\end{equation}
where $L(M)$ is defined in Equation~\ref{ssrs2mr:eqL}.
In order to recover the correct pair statistics, each companion 
must be assigned weights $w_{N2}(z_i)=1/S_N(z_i)$ and 
$w_{L2}(z_i)=1/S_L(z_i)$.
The total number and luminosity of close companions for the 
$i^{th}$ primary galaxy, computed by summing over the $j$ 
galaxies satisfying the ``close companion'' criteria, 
is given by 
$N_{c_i} = \sum_j{w_{N2}(z_j)}$ and
$L_{c_i} = \sum_j{w_{L2}(z_j)L_j}$ respectively.
By applying this weighting scheme to all galaxies in the 
secondary sample, we will retrieve pair statistics that 
correspond to a volume-limited sample with $\mbright \leq M \leq M_2$.  

\subsubsection{Weighting of Primary Sample} \label{ssrs2mr:w1}

The above weighting scheme ensures that the number and luminosity 
of companions 
found around each primary galaxy is normalized 
to $\mbright \leq M \leq M_2$.  
However, these estimates are obviously better for galaxies at 
the low redshift end of the primary sample, since they will have 
the largest number of {\it observed} companions.  Recall that 
$N_c$ and $L_c$ are quantities that are averaged over a sample 
of primary galaxies.  In order to minimize the errors in these 
statistics, we assign weights to the primary galaxies 
(denoted $w_{N1}(z_i)$ and 
$w_{L1}(z_i)$) which are inversely 
proportional to the square of their uncertainty.  
If the observed number and luminosity of companions 
around the $i^{th}$ primary 
galaxy are given by $N_i(obs)$ and $L_i(obs)$ respectively, 
and if we assume that the uncertainties are determined by 
Poisson counting statistics, then 
$N_{c_i} = w_{N2}(z_i) N_i(obs) \pm w_{N2}(z_i) \sqrt{N_i(obs)}$ 
and 
$L_{c_i} = w_{L2}(z_i) L_i(obs) \pm w_{L2}(z_i) \sqrt{L_i(obs)}$. 
On average, these quantities will be related to 
expectation values $<N_c>$ and $<L_c>$ by 
$<N_c> = w_{N2}(z_i) N_i(obs)$ and 
$<L_c> = w_{L2}(z_i) L_i(obs)$.  Combining these relations yields 
\begin{equation} \label{ssrs2mr:eqndelN1}
w_{N1}(z_i) = {1 \over N_i(obs) w_{N2}(z_i)^2} = {1 \over <N_c>
w_{N2}(z_i)} \propto w_{N2}(z_i)^{-1}
\end{equation}

\begin{equation} \label{ssrs2mr:eqndelL1}
w_{L1}(z_i) = {1 \over L_i(obs) w_{L2}(z_i)^2} = {1 \over <L_c>
w_{L2}(z_i)} \propto w_{L2}(z_i)^{-1}
\end{equation}

That is, the optimal weighting is the 
reciprocal of the weighting scheme used for companions.  
Therefore, weights 
$w_{N1}(z_i)=S_N(z_i)$ and $w_{L1}(z_i)=S_L(z_i)$ should be 
assigned to primary galaxies.
The pair statistics are then computed as follows :
\begin{equation} \label{ssrs2mr:eqNcobs}
{N_c} = {\sum_i w_{N1}(z_i)N_{c_i} \over \sum_i w_{N1}(z_i)}
\end{equation}
\begin{equation} \label{ssrs2mr:eqLcobs}
{L_c} = {\sum_i w_{L1}(z_i)L_{c_i} \over \sum_i w_{L1}(z_i)} .
\end{equation}
It is worth noting that, for a close pair, both galaxies 
will lie at roughly the same redshift, meaning that 
$w_{1}(z_i)\times w_{2}(z_j) \approx 1$.  We choose not to 
make this approximation, in order to keep these relations 
valid for pairs that are not close, and to allow for 
future application to pairs with additional selection weights.  
However, we stress that, with or without this approximation, the 
primary weights in the denominator provide an overall correction 
for the flux limit, unlike the traditional pair fraction. 
Note also that, for a volume-limited sample, weights for 
all galaxies in the primary and secondary samples are equal, 
reducing these equations to 
${N_c} = \sum N_{c_i} / N_1$ and 
${L_c} = \sum L_{c_i} / N_1$, 
as defined in Section~\ref{ssrs2mr:realspace}.

\subsubsection{Boundary Effects}\label{ssrs2mr:wb}
A small correction must be made to these weights 
if a primary galaxy 
lies close to a region of space that is not covered by the 
survey.  
This will happen if a galaxy lies close to the boundaries 
on the sky, or close to the minimum or maximum redshift 
allowed.  
If this is the case, it is possible that close 
companions will be missed, leading to an underestimate of 
the pair statistics.  
Therefore, we must account for these effects.  

First, we consider galaxies lying close to the survey 
boundaries on the sky, as defined in Section~\ref{ssrs2mr:data}.  
For each galaxy in the primary sample, we compute the fraction 
of sky with $\rpmin \leq r_p \leq \rpmax$ that lies within 
the survey boundaries.  
This fraction will be denoted $\fb$.  
For SSRS2, our usual choices of $\rpmin$ and $\rpmax$ (see 
\S~\ref{ssrs2mr:class}) make this a very small effect, 
with $f_b$=1 for 99.75\% of galaxies in the primary sample.   
Having measured $\fb$ for each galaxy in the primary 
sample, we must incorporate this into the measurement of the 
pair statistics.  The first task is to ensure that we correct 
the number of companions to match what would be expected if 
coverage was complete.  We do this by assigning each companion 
a boundary weight $w_{b_2}$ = $1/\fb$, where $\fb$ is associated with 
its host galaxy from the primary sample.  
By multiplying each companion by 
its boundary weight, we will recover the correct number of 
companions.  We must also adjust weights for the primary galaxies.  
Following the method described in the previous section, 
we wish to give less weight to galaxies that 
are likely to have fewer observed companions.  
Therefore, each 
primary galaxy is assigned a boundary weight $w_{b_1}$ = $\fb$.  

We now consider galaxies which lie near the survey 
boundaries along the line of sight.  If a primary galaxy lies 
close to the minimum or maximum redshift allowed, it is 
possible that we will miss companions because they lie just 
across this redshift boundary.  In order to 
account correctly 
for this effect, one would need to model the velocity 
distribution of companions.  As this requires several 
assumptions, we choose instead to exclude all companions 
that lie between a primary galaxy and its nearest 
redshift boundary, provided the boundary lies within 
$\Delta \vmax$ of the primary galaxy.  
To account for this exclusion, 
we assume that the velocity distribution is symmetric along 
the line of sight.  Thus, as we will miss half of the companions 
for these galaxies, we assign a weight of 
$w_{2_v}$=2 to any companions found in the direction opposite 
to the boundary.  We must also consider how to weight the 
primary galaxies themselves.  Clearly, primary galaxies close 
to the redshift boundaries will be expected to have half as 
many {\it observed} companions as other primary galaxies.  
To minimize the errors in computing the pair statistics, 
we assign these primary galaxies weights $w_{v_1}$=0.5.  

To summarize, weights for companions in the secondary sample 
are given by 
\begin{equation}\label{ssrs2mr:eqnwn2}
w_{N_2}=S_N(z_i)^{-1} w_{b_2}w_{v_2},
\end{equation}
\begin{equation}\label{ssrs2mr:eqnwl2}
w_{L_2}=S_L(z_i)^{-1} w_{b_2}w_{v_2},
\end{equation}
while primary galaxies are assigned weights 
\begin{equation}\label{ssrs2mr:eqnwn1}
w_{N_1}=S_N(z_i) w_{b_1}w_{v_1},
\end{equation}
\begin{equation}\label{ssrs2mr:eqnwl1}
w_{L_1}=S_L(z_i) w_{b_1}w_{v_1}.
\end{equation}

\subsection{Confirmation Using Monte Carlo Simulations}
We will now perform a test to see if this weighting scheme 
achieves the desired effects.  
To do this, we will use flux-limited 
Monte Carlo simulations, 
for which the intrinsic density and clustering are fixed.  
Therefore, the {\it intrinsic} pair statistics do not depend
on redshift or luminosity.  If the secondary sample 
weights are correct, 
the measured pair statistics will be the same everywhere 
(within the measurement errors), regardless of redshift or 
luminosity.  We will also check to see if 
the weights for the primary sample are correct.  If they are, 
the errors on the pair statistics will be minimized, as 
desired.

The flux-limited Monte Carlo simulations 
were generated in a similar manner to the simulations 
described in Section~\ref{ssrs2mr:MC}; however, 
a limiting apparent magnitude of $m_B \leq 15.5$ 
was imposed.  Sample sizes of 8000 (South) and 
4035 (North) were used, providing a good match to the 
overall density in SSRS2.  
The resulting simulations are similar 
to SSRS2 in all respects, except for the absence of clustering.  
We have already established how the pair statistics depend on 
the choice of $M_1$ and $M_2$.  In the following analysis, 
we choose $M_2$=$M_1$=$-19$.  
In Section~\ref{ssrs2mr:clust}, we outlined reasons for 
restricting the sample to a fixed range in absolute magnitude.  
Here, we demonstrate how the chosen range affects $N_c$ and 
$L_c$.  For comparison, we also compute $N_c$ without normalizing 
to a specified range in absolute magnitude (in this case,  
$w_{N_1}$=$w_{N_1}$=1).  This provides some 
insight into the behaviour of the traditional (uncorrected) 
pair fraction.

These tests are most straightforward if the intrinsic pair 
statistics are the same everywhere in the enclosed volume.  This 
is not quite true for these simulations, however.  Galaxies
are distributed randomly within the enclosed {\it co-moving} 
volume.  As a result, the physical density varies with redshift 
as $(1+z)^3$.  In addition, the volume element 
encompassed by the line-of-sight pair criterion $\Delta v$ 
varies with redshift as $(1+z)^{-3/2}$ for $q_0$=0.5.  
In order to have the simulations mimic a sample with universal 
pair statistics, 
we normalize the sample for these effects by weighting 
each galaxy by $(1+z)^{-3/2}$.  We stress that this is done 
only for the Monte Carlo simulations.  
One should {\em not} apply either of these corrections to real 
redshift data.

\begin{figure}
\unitlength1.0cm
\begin{center}
\begin{picture}(16.0,16.0)
\put(0,0)
{\epsfig{file=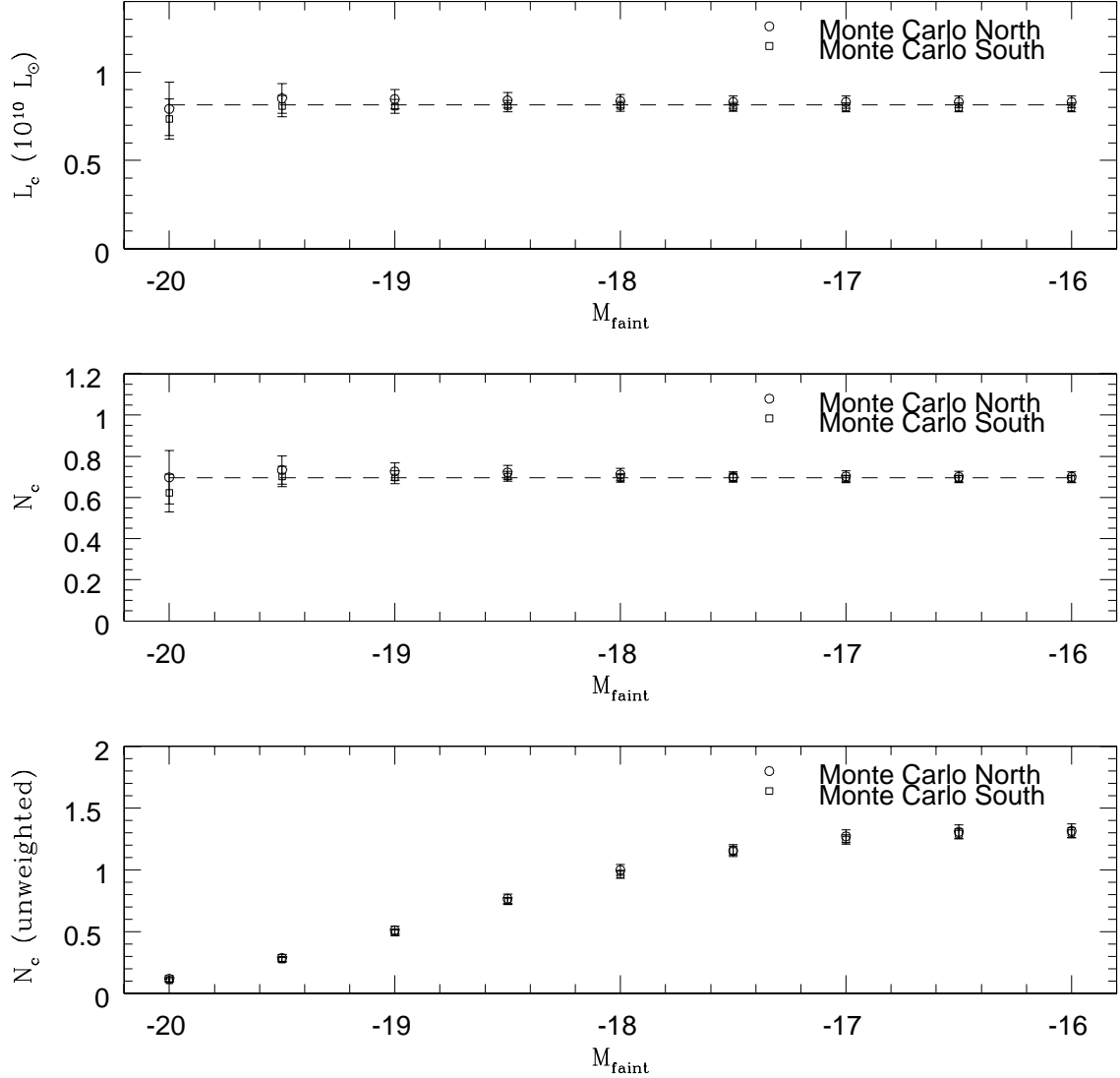,height=16.0cm,width=16.0cm}}
\end{picture}
\end{center}
\figcaption[figrmfaint.ps]
{
Pair statistics are computed for the flux-limited Monte Carlo 
simulations.  Three pair statistics (unweighted $N_c$, $N_c$, and $L_c$) 
are given, for a range of choices 
of minimum luminosity $\mfaint$, with $M_2$=$M_1$=$-19$.  
Error bars are computed using the Jackknife technique.
The unweighted $N_c$ exhibits a systematic 
increase with $\mfaint$.  This is a consequence of not correcting for 
the corresponding increase in mean galaxy number density.  
This bias is taken into account when computing $N_c$ and $L_c$. 
The horizontal dashed lines match the data at $M_{faint}$=$-16$, 
and demonstrate that $N_c$ and $L_c$ do not depend on 
$M_{faint}$.
\label{ssrs2mr:figrmfaint}}
\end{figure}

In Figure~\ref{ssrs2mr:figrmfaint}, the pair statistics are computed for 
a range of $\mfaint$.  In addition, we compute $N_c$ without weighting 
by the luminosity function, to demonstrate the danger of ignoring 
this important correction.  This statistic is directly analagous to 
the traditional (uncorrected) pair fraction used in the literature.  
It is clear that both $N_c$ and $L_c$ 
are independent of the choice of $\mfaint$, within the errors.  This 
verifies that we have correctly accounted for the biases 
introduced by the apparent magnitude limit.  In contrast, 
the unweighted $N_c$ is seen to have a strong dependence on $\mfaint$.  As 
expected, it increases as $\mfaint$ becomes fainter, 
due to the increase in sample density.  
We stress that this 
does not happen with the normalized $N_c$ and $L_c$ statistics,  
because both 
are corrected to a fixed range in limiting absolute magnitude.  

\begin{figure}
\unitlength1.0cm
\begin{center}
\begin{picture}(16.0,16.0)
\put(0,0)
{\epsfig{file=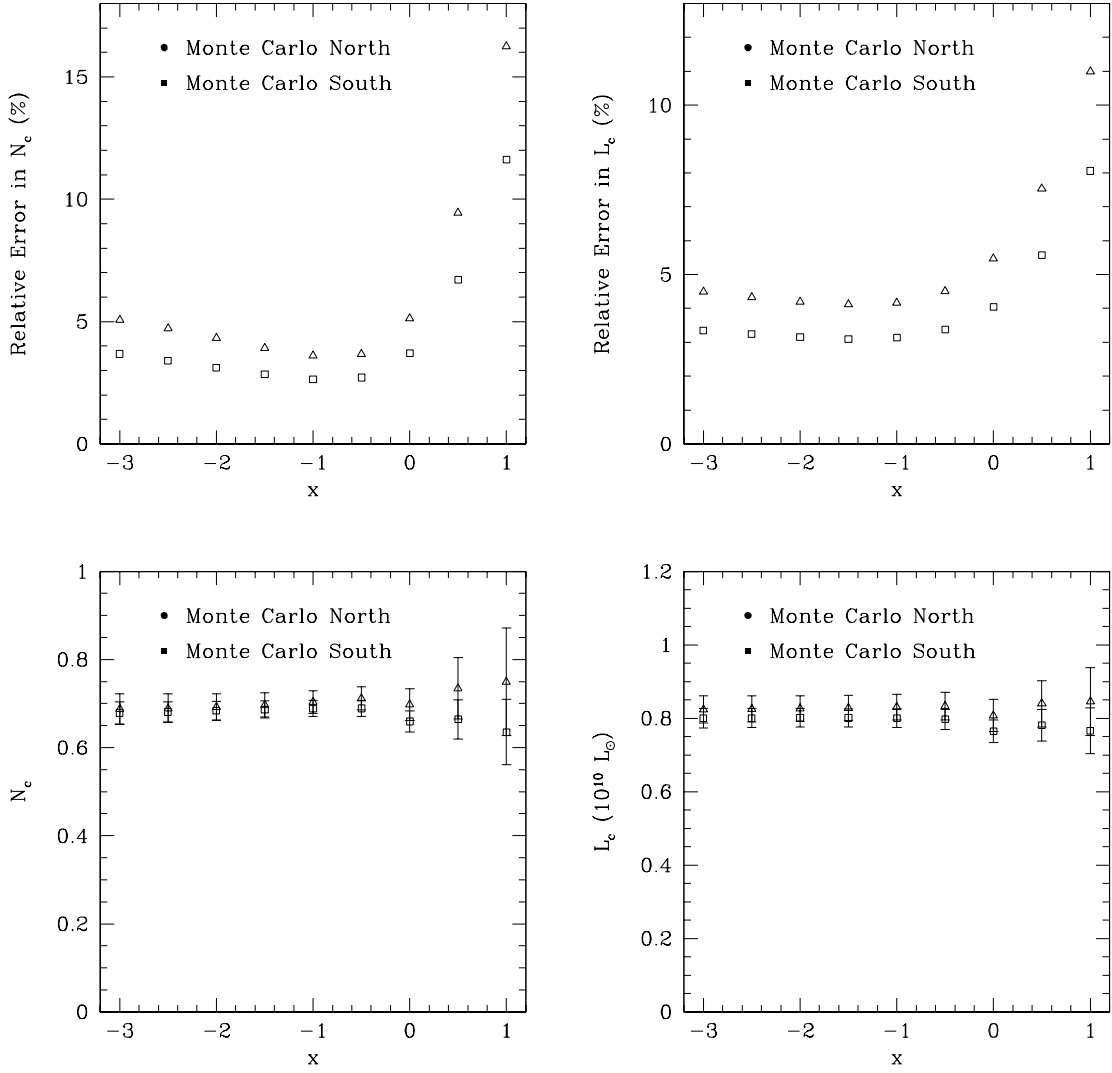,height=16.0cm,width=16.0cm}}
\end{picture}
\end{center}
\figcaption[figw1.ps]
{
Pair statistics are computed for a range of possible weighting 
schemes for primary galaxies.  
Error bars are computed using the Jackknife technique.
\label{ssrs2mr:figw1}}
\end{figure}

Finally, we demonstrate that the weighting scheme 
used for the primary sample (\S~\ref{ssrs2mr:w1}) does in fact 
minimize errors in $N_c$ and $L_c$.  
Recall that the weighting used was the reciprocal of the 
weights for the secondary sample.  
Here we will assume that 
$w_{N1} \propto w_{N2}(z_i)^{x}$ and 
$w_{L1} \propto w_{L2}(z_i)^{x}$.  In Section~\ref{ssrs2mr:w1}, 
justification was given for setting $x$=$-1$.  Here, we will 
allow $x$ to vary, in order to investigate empirically which 
value minimizes the errors.  Special cases of interest are 
$x$=0 (no weighting) and $x$=1 (same weighting as  
{\it secondary} sample).  
The results are given in Figure~\ref{ssrs2mr:figw1}.  
The relative errors in $N_c$ and $L_c$ reach a minimum at 
$x~\sim -1$, as expected.  
Errors are $\sim$ 40\% larger if no weighting is used ($x$=0).  
For $x$=1, errors are much larger, increasing by nearly a factor 
of 5.  While errors increase dramatically for $x \ge 0$, they 
change slowly around $x$=$-1$.  
Clearly, $x$ =$-1$ is an excellent choice.

\section{A PAIR CLASSIFICATION EXPERIMENT} \label{ssrs2mr:class}
The first step in applying these techniques to a real survey 
of galaxies is to decide on a useful close pair definition.  
This involves imposing a maximum 
projected physical separation ($\rpmax$) and, if possible,  
a maximum line-of-sight rest-frame velocity difference 
($\Delta \vmax$).  
The limits should be chosen so as to extract information 
on mergers in an optimal manner.  
This involves a compromise between the number and merging 
likelihood of pairs.  
While one should focus on companions which are most 
likely to be involved in mergers,   
a very stringent pair definition may yield a small and 
statistically insignificant sample.  

\begin{figure}
\unitlength1.0cm
\begin{center}
\begin{picture}(16.0,16.0)
\put(0,0)
{\epsfig{file=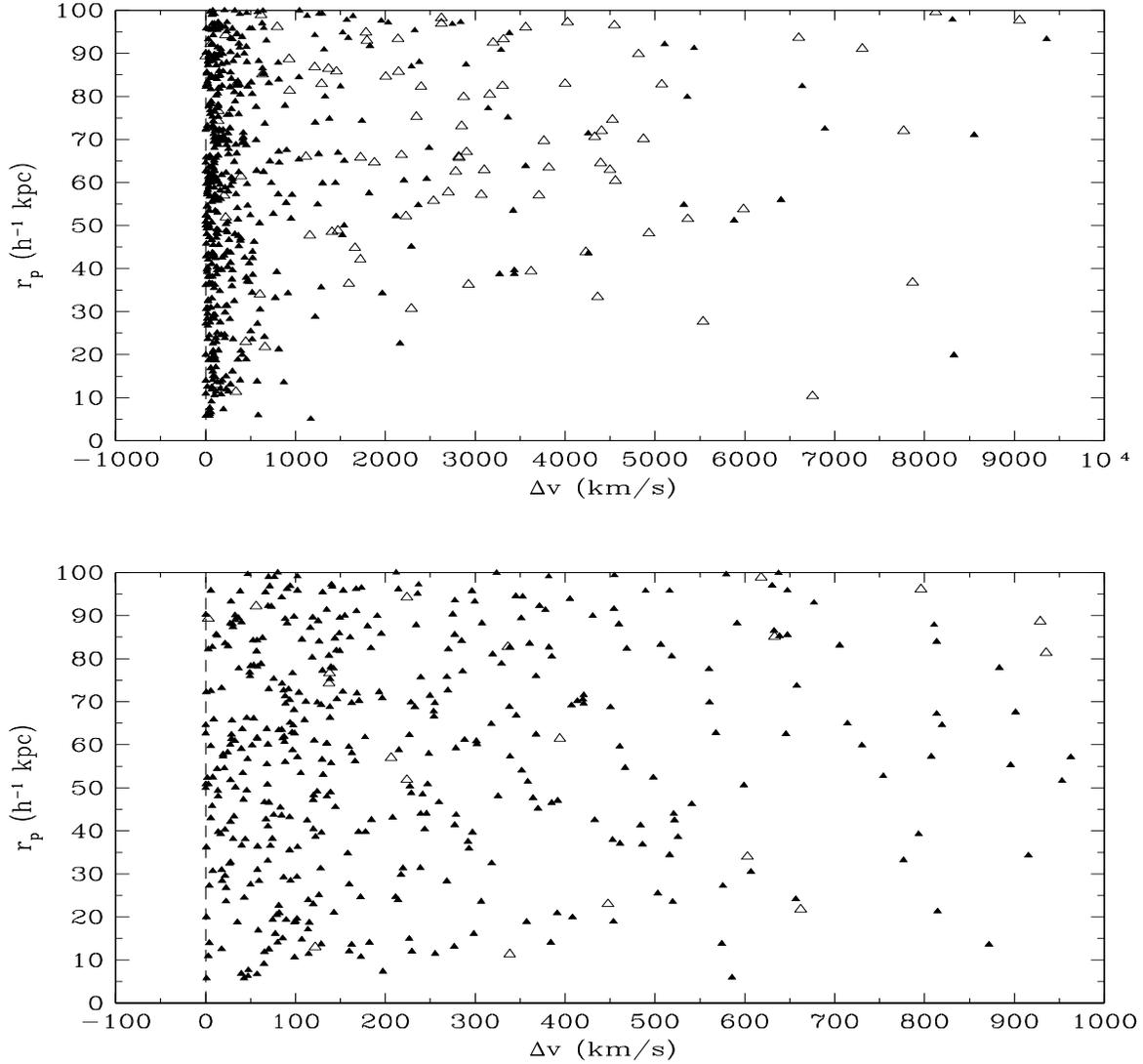,height=16.0cm,width=16.0cm}}
\end{picture}
\end{center}
\figcaption[figrpdelv.ps]
{
Projected separation ($r_p$) is plotted versus rest-frame velocity
difference ($\Delta v$) for all pairs with 
$r_p \leq$ 100 \hkpc.  
Pairs were identified by looking for close companions (real or 
random) around primary galaxies in the SSRS2 survey.  
Filled triangles represent real companions found in the SSRS2 
survey.  Open triangles denote random companions found in the 
Monte Carlo simulations.  Note that the density of the simulations 
was matched to the real data sample when generating this plot, 
ensuring a fair comparison.  
The upper plot gives a wide range in velocity differences, 
demonstrating how the number of companions in excess of random 
becomes negligible beyond $\sim$ 1000 km/s.  The lower plot 
is a zoomed-in version 
of the upper plot, giving a better feel for the distribution of 
physical pairs.  
\label{ssrs2mr:figrpdelv}}
\end{figure}

In previous close pair studies, 
the convention has been to set $\rpmax$ = 20 \hkpc.  
Pairs with separations of $r_p \leq$ 20 \hkpc~ are 
expected to merge within 0.5 Gyr (e.g., \cite{B88}, \cite{P97}).  
We note, however, that timescale estimates are approximate in 
nature, and have yet to be verified.
In earlier work, it has not been possible to apply a velocity 
criterion, since redshift samples have been too small to 
yield useful pair statistics using only galaxies 
with measured redshifts.  Instead, all physical companions 
have been used, with statistical correction for optical 
contamination (\cite{P97}).  With a complete redshift 
sample, we can improve on this.  
This can be seen by inspecting a plot of $r_p$ versus $\Delta v$ 
for the SSRS2 pairs, given in Figure~\ref{ssrs2mr:figrpdelv}.  
By imposing a velocity 
criterion, we can eliminate optical contamination; furthermore, 
we are able to concentrate on the physical pairs with the 
lowest relative velocities, and hence the greatest likelihood 
of merging.  

\begin{figure}
\unitlength1.0cm
\begin{center}
\begin{picture}(16.0,16.0)
\put(0,0)
{\epsfig{file=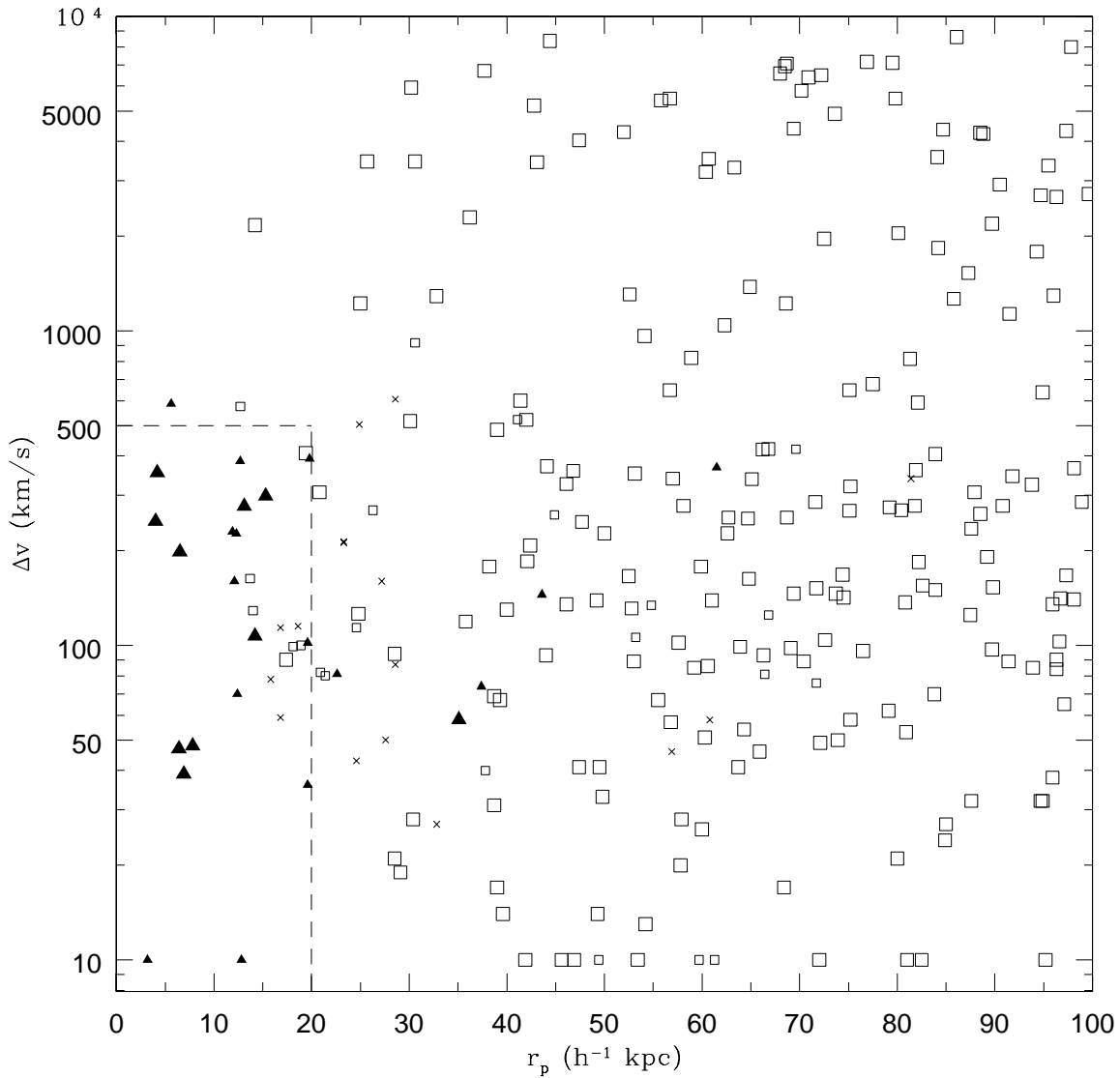,height=16.0cm,width=16.0cm}}
\end{picture}
\end{center}
\figcaption[figpc.ps]
{
Pair interaction classification (\Ic) is plotted as a function 
of projected physical separation ($r_p$) and line-of-sight 
rest-frame velocity difference ($\Delta v$), 
for 255 unique SSRS2 pairs with 
$r_p <$ 100 \hkpc.  
Symbols are defined as follows~:
\Ic=0,1 (large open squares), 
\Ic=2,3 (small open squares),
\Ic=4$-$6 (crosses),
\Ic=7,8 (small filled triangles),
and
\Ic=9,10 (large filled triangles).
High \Ic~indicates increased likelihood of interaction.
For plotting convenience, pairs with $\Delta v < 10$ km/s 
are assigned $\Delta v = 10$ km/s.  
The dashed line marks the close pair criteria 
($r_p$=20~\hkpc~ and $\Delta v$=500 km/s) used for 
computing pair statistics.  
\label{ssrs2mr:figpc}}
\end{figure}

We can now use our large sample of low-$z$ pairs to 
shed new light on these issues.  We will use images of 
these pairs in an attempt to determine how signs of 
interactions are related to pair separation.  
We begin by finding all 255 SSRS2 pairs with 
$r_p \leq 100$~\hkpc,   
computing $r_p$ and $\Delta v$ for each.
Images for these pairs 
were extracted from the Digitized Sky Survey.  
Interactions were immediately apparent in
some of these pairs, and the images were deemed to 
be of sufficient quality that a visual classification 
scheme would be useful.  
An interaction classification parameter (\Ic) was devised, 
where \Ic=0 indicates that a given pair is 
``definitely not interacting'', and \Ic=10 
indicates ``definitely interacting''.  
In order to avoid a built-in bias, the classifier is not 
given the computed values of $r_p$ and $\Delta v$.  
The classifier uses all visible information available 
(tidal tails and
bridges, distortions/asymmetries in member galaxies, 
apparent proximity, etc.).  
Classifications were performed by three of us (DRP, ROM, RGC), 
and the median classification was determined for each system. 
The results are presented in Figure~\ref{ssrs2mr:figpc}.  
A clickable version of this plot, which allows the user to 
see the corresponding Digitized Sky Survey image for each pair, 
is available at http://www.astro.utoronto.ca/$\sim$patton/ssrs2/Ic.  

There are several important features in this plot.  First, 
there is a clear correlation between $I_c$ and $r_p$, with 
closer pairs exhibiting stronger signs of interactions.  
There are several interacting pairs with $r_p \sim 50$ \hkpc.  
While these separations are fairly large, 
it is not surprising that 
there would be some early-stage mergers with these separations  
(e.g., Barton, Bromley, \& Geller 1998\nocite{BARTON98}).  
An excellent example of this phenomenon 
is the striking tail-bridge system Arp 295a/b 
(cf. \cite{HG96}), which has $r_p$ = 95~\hkpc.    
However, these systems clearly do not dominate; almost all 
pairs with large separations have very low interaction 
classifications.  
The majority of pairs showing clear signs of interactions/mergers 
have $r_p \lesssim 20$ \hkpc~.  

There is also a clear connection with $\Delta v$.  
Pairs with $\Delta v > 600$ km/s do not exhibit signs 
of interactions, with 61/63 (97\%) classified  
as $I_c \leq 1$.  This indicates that interactions are 
most likely to be seen in low velocity pairs, as expected.  
We note, however, that there are very few optical pairs 
(i.e., small $r_p$ and large $\Delta v$) in this low redshift 
sample.  At higher redshift, increased optical contamination 
may lead to difficulties in identifying interacting systems 
when the galaxies are close enough to have overlapping 
isophotes.  Clearly, it is necessary to have redshift 
information for both members of each pair if one is to 
exclude these close optical pairs.

After close inspection of Figure~\ref{ssrs2mr:figpc}, we decided on 
close pair criteria of 
$\rpmax$ = 20~\hkpc~ and $\Delta \vmax$ = 500 km/s.  
A mosaic of some of these pairs 
is given in Figure~\ref{ssrs2mr:figim}.  
In this regime, 31\% (9/29) exhibit convincing
evidence for interactions (\Ic~$ \geq 9$), while 69\% (20/29) 
show some indication of interactions (\Ic~$\geq 6$).  
Furthermore, the vast majority (9/10) of pairs with clear 
signs of interactions (\Ic~$ \geq 9$) are found in this regime.   
These criteria appear to identify a sample of pairs which are 
likely to be undergoing mergers; moreover, the resulting 
sample includes most of the systems classified as interacting.  

We also impose an inner boundary of $r_p$ = 5 \hkpc.  
This limit is chosen so as to avoid the confusion that is often 
present on the smallest scales.  
In this regime, it is often difficult to distinguish between 
small galaxies and sub-galactic units, particularly 
in merging systems.  
While we are omitting the most likely merger
candidates, those at separations $<$~5~\hkpc~ are not expected
to account for more than $\sim$ 5\% of the companions 
within 20 \hkpc~.  This expectation, which has yet to be 
verified, is based both on pair counts in {\it HST} imaging 
(\cite{BKWF}) and on inward extrapolation of the 
correlation function (\cite{P97}).   
While this inner boundary will lead to a slight decrease in $N_c$ 
and $L_c$, it should have no significant effect on estimates of 
merger/accretion rate evolution, provided the same restriction 
is applied to comparison samples at other redshifts.  

\section{APPLICATION TO SSRS2} \label{ssrs2mr:sample}
In the preceding sections, we have outlined techniques for 
measuring pair statistics in a wide variety of samples.  
We have demonstrated a robust method of applying this approach 
to flux-limited samples, accounting for redshift-dependent 
density changes and 
minimizing differences in clustering.  We have also selected 
pair definitions that identify the most probable imminent mergers.  
We will now apply these techniques to the SSRS2 survey. 
As this is a complete redshift survey, redshifts are available 
for all close companions; hence, for the first time, we will 
measure pair statistics using only close {\it dynamical} pairs. 
After limiting the analysis to a reasonable range in absolute 
magnitude, we compute $N_c$ and $L_c$ for the SSRS2 survey.

\subsection{Defining Survey Parameters}
\begin{figure}
\unitlength1.0cm
\begin{center}
\begin{picture}(16.0,16.0)
\put(0,0)
{\epsfig{file=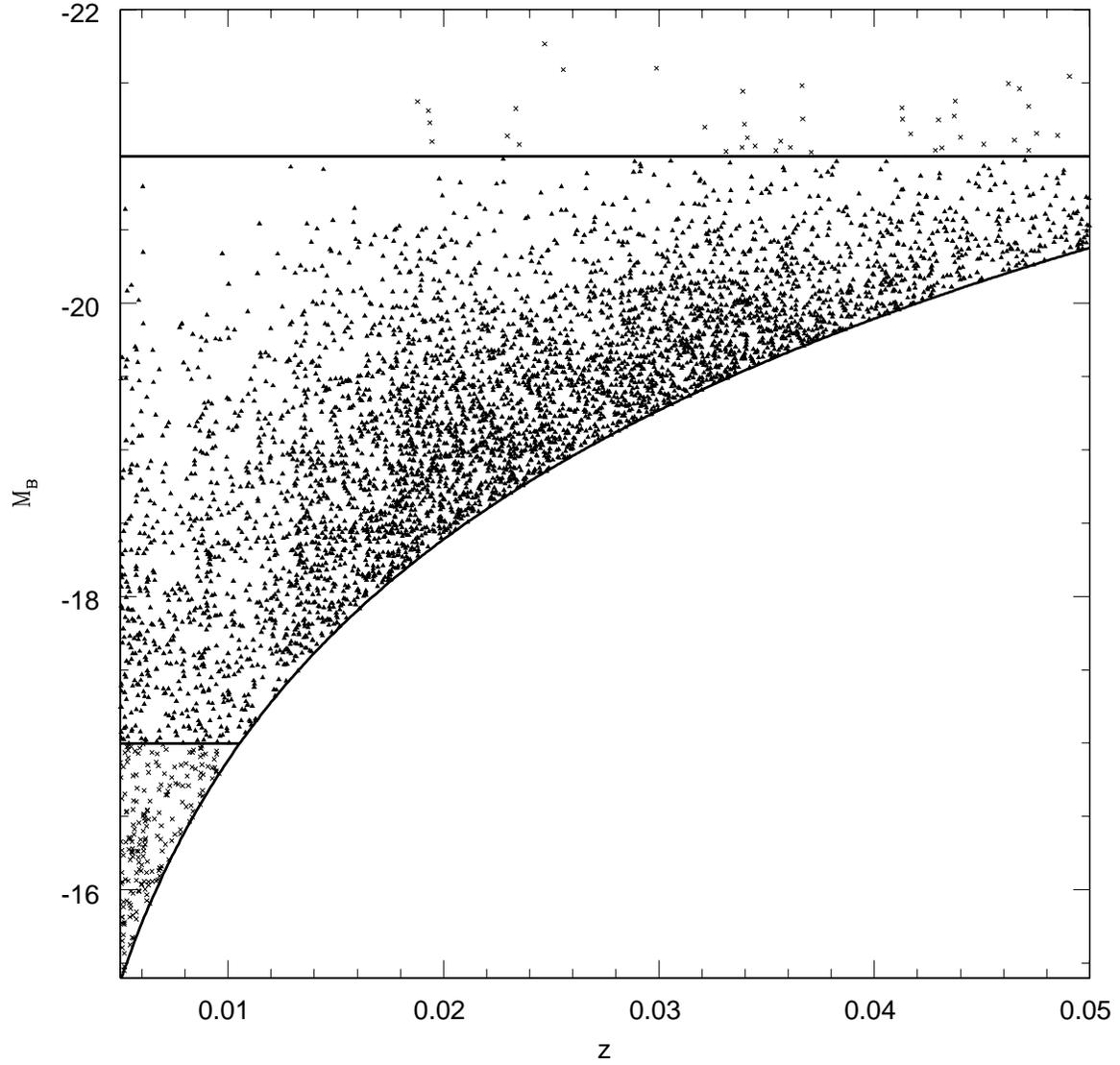,height=16.0cm,width=16.0cm}}
\end{picture}
\end{center}
\figcaption[figmbz.ps]
{$B-$band absolute magnitude is plotted versus redshift 
for all SSRS2 galaxies with $m_B \leq 15.5$ and 
$0.005 \leq z \leq 0.05$.  
The curved line marks the boundary imposed by this apparent 
magnitude limit.  
The upper horizontal line indicates the bright limit imposed 
($\mbright$ = $-21$), while the lower horizontal line denotes 
the minimum luminosity allowed ($\mfaint$ = $-17$).  
Galaxies satisfying all of these criteria (and hence used 
in the calculation of $N_c$ and $L_c$) are marked 
with triangles; the remainder are indicated with crosses.  
\label{ssrs2mr:figmbz}}
\end{figure}

In Section~\ref{ssrs2mr:clust}, we emphasized the importance of 
restricting the sample in absolute magnitude, 
to minimize bias due to luminosity-dependent 
clustering.  For SSRS2, we first impose a bright limit of 
$\mbright$ = $-21$.  All galaxies brighter than this are hereafter 
excluded from the analysis.  This allows us to avoid the most 
luminous galaxies, which are probably the most susceptible to 
luminosity-dependent clustering; however, this reduces the size 
of the sample by only 0.5\%.  
We also impose a faint absolute magnitude limit of 
$\mfaint$ = $-17$, which results 
in the exclusion of intrinsically faint galaxies 
at $z \lesssim 0.01$.  This guards against the possibility that these 
intrinsically faint galaxies are clustered differently than the bulk 
of the galaxies in the sample.  
This pruning of the sample is illustrated in Figure~\ref{ssrs2mr:figmbz}.  
These restrictions allow us to minimize concerns about
luminosity-dependent clustering while retaining 90\% of the sample.  
The final results are insensitive to 
these particular choices (see Section~\ref{ssrs2mr:sense}).  

The above limits in absolute magnitude, along with the flux limit, 
define the usable sample of galaxies.  In order to compute pair 
statistics, we must also normalize the measurements to a given 
range in absolute magnitude, for both the primary and 
secondary samples.  
The mean limiting absolute magnitude of the primary 
sample, weighted according to Section~\ref{ssrs2mr:w1}, 
is $M_B$ = $-17.8$.  For convenience, 
we set $M_2(B)$ = $-18.0$ (we will compute pair statistics for 
$-19 \leq M_2(B) \leq -17$ in the following section).  
For reference, we note that this corresponds to 
$m_B$=$15.5$ at $z$=0.017.  
As we are dealing with 
a complete redshift sample, we set $M_1(B)$=$M_2(B)$ in order to 
use all of the available information.
Finally, as we have limited the sample using $\mbright$ = $-21$, 
this will be used in conjunction with $M_2(B)$ to derive 
pair statistics for galaxies with $-21 \leq M_B \leq -18$.  

\subsection{SSRS2 Pair Statistics} \label{ssrs2mr:ssrs2}

\begin{figure}
\unitlength1.0cm
\begin{center}
\begin{picture}(16.0,16.0)
\put(0,0)
{\epsfig{file=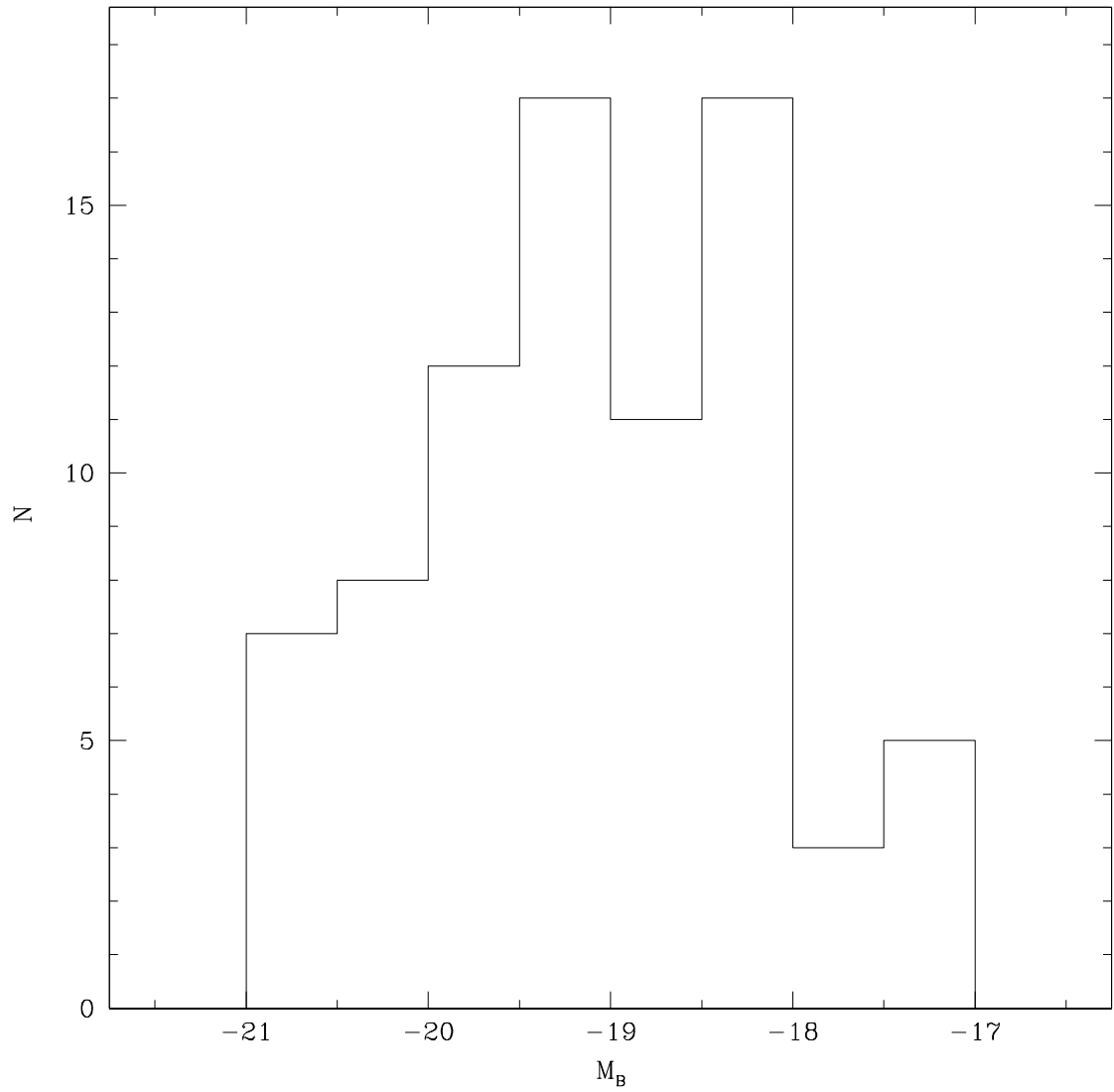,height=16.0cm,width=16.0cm}}
\end{picture}
\end{center}
\figcaption[figlh.ps]
{An absolute magnitude histogram is given for the 80 companions 
used in the pair statistics.  
\label{ssrs2mr:figlh}}
\end{figure}

Using these parameters, we identified all close 
companions in SSRS2.
The North sample yielded 27 companions, and 53 were found in 
the South, giving a total of 80.  We emphasize that it is 
{\it companions} that are counted, rather than pairs; hence, 
if both members of a pair fall within the primary sample, 
the pair will usually yield 2 companions.    
A histogram of companion absolute magnitudes is given in 
Figure~\ref{ssrs2mr:figlh}.  This plot shows that 90\% of 
the companions we observe in our flux-limited sample 
fall in the range $-21 \leq M_B \leq -18$.  Hence, galaxies
with $-18 \leq M_B \leq -17$ do not dominate the sample.
Tables~\ref{ssrs2mr:tabcpn} and~\ref{ssrs2mr:tabcps} give complete 
lists of close aggregates (pairs and triples) for SSRS2 
North and South respectively.
These systems contain all companions used in the computation of 
pair statistics. 
These tables list system ID, number of members, $r_p$ (\hkpc), 
$\Delta v$ (km/s), RA (1950.0), DEC (1950.0), and recession 
velocity (km/s).  
DSS images for these systems were given earlier in 
Figure~\ref{ssrs2mr:figim}. 

Using this sample of companions, the pair statistics 
were computed.  
The results are given in Table~\ref{ssrs2mr:tabstats}.  
Errors were computed using the Jackknife technique.  
Results from the two subsamples were combined, 
weighting by Jackknife errors, to give $N_c = 0.0226 \pm 0.0052$ 
and $L_c = 0.0216\pm 0.0055 \times 10^{10}~h^2 \lsun$ 
at $z$ = 0.015.  Results from the two subsamples agree within 
the quoted 1$\sigma$ errors.  

To facilitate future comparison with other samples, we also 
generate pair statistics spanning the range 
$-19 \leq M_2(B) \leq -17$ (see Table~\ref{ssrs2mr:tabm2}).  
We note, however, that while we account for changes in 
number and luminosity density 
over this luminosity range (using LF weights described in 
Section~\ref{ssrs2mr:flux}), there is no correction for changes in 
clustering.  
Hence, our statistics should be considered most appropriate 
for $M_2(B)$ = $-18$, and more approximate in nature at brighter and 
fainter levels.  
The results in Table~\ref{ssrs2mr:tabm2} indicate that $N_c$
increases by a factor of 5 between 
$M_2(B)$ = $-19$ and $M_2(B)$ = $-17$, resulting solely from 
an increase in 
mean number density.  The change in $L_c$ is less pronounced, with 
an increase by a factor of 2 over the same luminosity range.  
These substantial changes in both statistics emphasize the 
need to specify $M_2$ when computing pair statistics and comparing 
results from different samples.  
In addition, 
the smaller change in $L_c$ is indicative of 
the benefits of using a luminosity 
statistic such as $L_c$, which is more likely to converge 
as one goes to fainter luminosities (see Section ~\ref{ssrs2mr:MC}).  
$L_c$ will always converge faster than $N_c$, thereby reducing 
the sensitivity to $M_2$.  
Furthermore, it is possible to retrieve most of the relevant 
luminosity information without probing to extremely faint 
levels.
For example, for the SSRS2 LF, 70\% of the 
total integrated luminosity density is sampled by probing down to 
$M_2(B)$ = $-18$.  To first order, the same will be true for $L_c$.  
Going 2 magnitudes fainter would increase the
completeness to 95\%.  While we are currently unable to apply pair 
statistics down to these faint limits, this will be 
pursued when deeper surveys become available.  

\subsection{Sensitivity of Results to Survey Parameters} 
\label{ssrs2mr:sense}

In this section, we explore the effects of choosing different 
survey parameters.  Earlier in this study, we demonstrated 
that $N_c$ and $L_c$ are insensitive to the choice of survey limits 
in absolute magnitude, provided clustering is independent 
of luminosity and the pair statistics are normalized correctly.  
Here, we test this hypothesis empirically.  

\begin{figure}
\unitlength1.0cm
\begin{center}
\begin{picture}(16.0,16.0)
\put(0,0)
{\epsfig{file=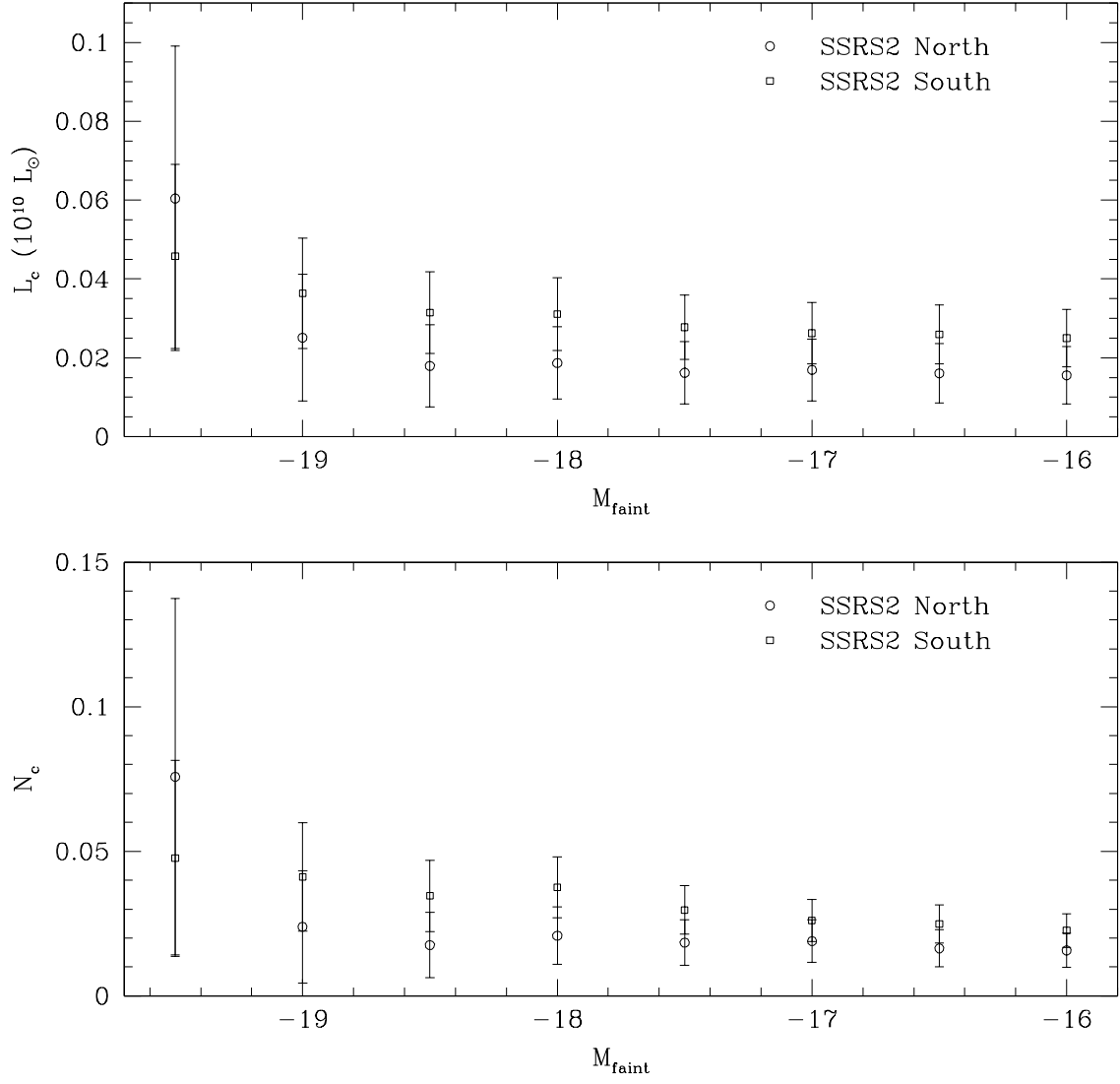,height=16.0cm,width=16.0cm}}
\end{picture}
\end{center}
\figcaption[figmfaint.ps]
{
Pair statistics are computed for SSRS2 North and South. 
$N_c$ and $L_c$ are given, for a range in
minimum luminosity $\mfaint$, with $M_2$=$M_1$=$-18$.  
Error bars are computed using the Jackknife technique.
Both $N_c$ and $L_c$ appear to be independent of $\mfaint$, 
within the errors, over the range $-18 \leq \mfaint \leq -16$.  
This implies that, to first 
order, clustering is independent of luminosity in this regime.    
\label{ssrs2mr:figmfaint}}
\end{figure}

First, we compute the pair statistics for a range in 
$\mfaint$, normalizing the statistics to $-21 \leq M_B \leq -18$ in 
each case.  Figure~\ref{ssrs2mr:figmfaint} demonstrates a possible 
trend of decreasing pair statistics with fainter $\mfaint$.  
This trend, however, is significant only for the brightest 
galaxies ($\mfaint \lesssim -19$).  
This is consistent with 
the findings of Willmer et al. (1998)\nocite{W98}, 
who measure an increase in clustering for bright galaxies in 
SSRS2, on scales of $r_p > 1$~\hmpc.  
For fainter $\mfaint$, 
there is no significant dependence.  The pair statistics 
vary by $\sim$ 5\% over the range 
$-17.5 \leq \mfaint \leq -16.5$, 
which is well within the error bars.  
Therefore, we conclude that our choice of $\mfaint$ = $-17$ has a 
negligible effect on $N_c$ and $L_c$.  
This implies that, to first order, clustering is independent 
of luminosity within this sample.  

\begin{figure}
\unitlength1.0cm
\begin{center}
\begin{picture}(16.0,16.0)
\put(0,0)
{\epsfig{file=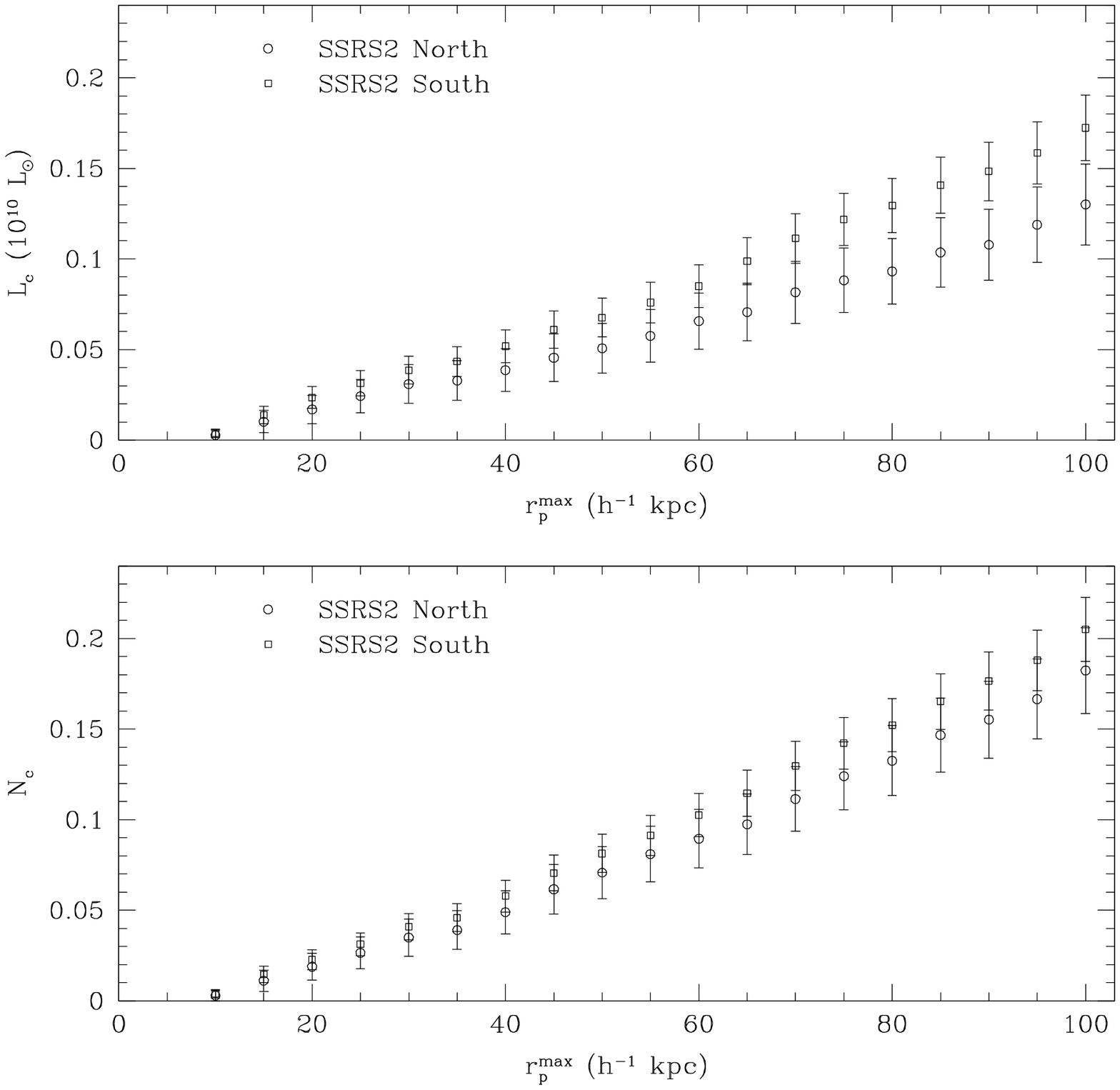,height=16.0cm,width=16.0cm}}
\end{picture}
\end{center}
\figcaption[figrp.ps]
{
Pair statistics are computed for $\Delta v \leq 500$ km/s, 
for a range of maximum projected separations ($\rpmax$).  
A minimum projected separation of $r_p=5$~\hkpc~is applied 
in each case.  
Error bars are computed using the Jackknife technique.
Both $N_C$ and $L_C$ are cumulative statistics; hence, 
measurements in successive bins are not independent.  
\label{ssrs2mr:figrp}}
\end{figure}

Next, we investigate how the pair statistics depend on 
our particular choices of $\rpmax$ and $\Delta \vmax$, 
which comprise our definition of a close companion.  
First, we compute pair statistics for 
10 \hkpc $\leq \rpmax \leq 100 $\hkpc, 
with $\Delta \vmax = 500$ km/s.  
Results are given in Figure~\ref{ssrs2mr:figrp}.  
This plot indicates a smooth increase in both statistics 
with $\rpmax$.  This trend is expected from measurements 
of the galaxy CF.  
The CF is commonly expressed as a power law of the 
form $\xi(r,z) = (r_0/r)^{\gamma}$, 
with $\gamma$=1.8  
(\cite{DP83}).
Integration over this function yields pair statistics that 
vary as $r_p^{3-\gamma} \approx r_p^{1.2}$, which is in 
good agreement with the trend found in Figure~\ref{ssrs2mr:figrp}.
From this plot, it also appears likely that there are systematic 
differences between the two subsamples.  
This is hardly surprising, since there are known differences 
in density between the subsamples, and it is likely that there 
are non-negligible differences in clustering as well.  This 
cosmic variance is not currently measurable on the smaller scales 
($r_p \leq 20$ \hkpc) relevant to our main pair statistics.  
Hence, we choose to ignore these differences for now.  However, 
these field-to-field variations are certain to add some systematic 
error to our quoted pair statistics.  

\begin{figure}
\unitlength1.0cm
\begin{center}
\begin{picture}(16.0,16.0)
\put(0,0)
{\epsfig{file=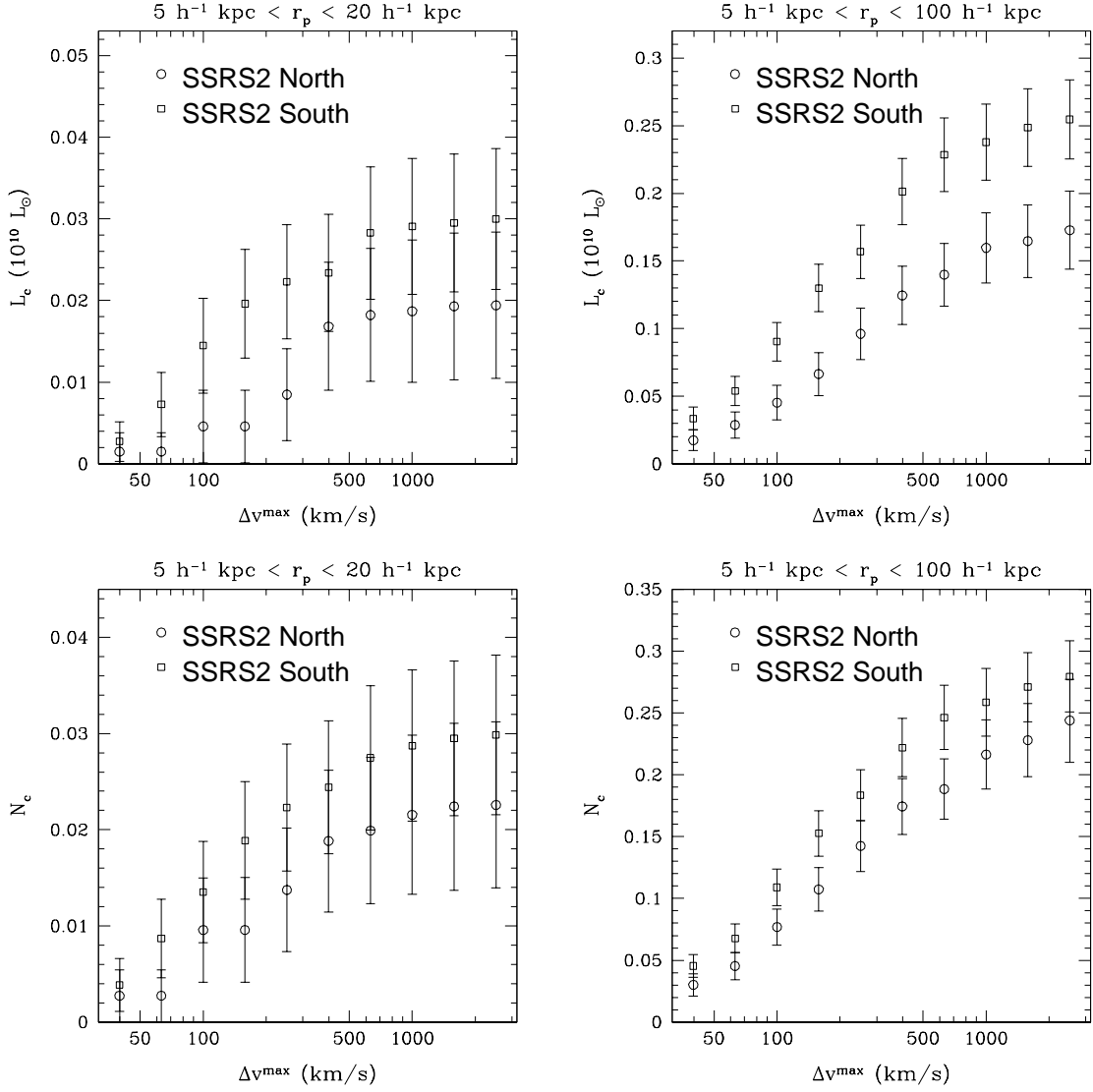,height=16.0cm,width=16.0cm}}
\end{picture}
\end{center}
\figcaption[figrl.ps]
{
Pair statistics are computed for $\Delta v \leq 500$ km/s, 
for a range of projected separations.  
Error bars are computed using the Jackknife technique.
Both $N_C$ and $L_C$ are cumulative statistics; hence, 
measurements in successive bins are not independent.  
\label{ssrs2mr:figrl}}
\end{figure}

We also compute pair statistics for a range in $\Delta \vmax$.  
This is done first for $\rpmax = 20 $~\hkpc, showing the 
relative contributions at different velocities to the main 
pair statistics quoted in this paper.  We also compute 
statistics using $\rpmax = 100~$\hkpc, in order to improve the 
statistics.  
Results are given in Figure~\ref{ssrs2mr:figrl}.  
Several important conclusions may be drawn 
from this plot.  
First, at small velocities ($\Delta \vmax \lesssim 700$ km/s), 
both pair statistics increase with $\Delta \vmax$, as 
expected.  This simply indicates that one continues to find 
additional companions as the velocity threshold increases.  
Secondly, it appears that our 
choice of $\Delta \vmax$ was a good one.  The $r_p \leq 20$ 
\hkpc~ pair statistics 
increase very little beyond $\Delta \vmax = 500$ km/s, 
while the contamination due to non$-$merging pairs would 
continue to increase (see Figure~\ref{ssrs2mr:figrpdelv}).   
Moreover, as both pair statistics flatten out at around 
$\Delta \vmax = 500$ km/s, small differences in the velocity 
distributions of different samples should not result in 
large differences in their pair statistics.
Finally, for $\rpmax \leq 100$ \hkpc, the pair 
statistics continue to increase out to $\sim 2000$ km/s.
This indicates an increase in velocity dispersion 
at these larger separations.  
This provides 
additional confirmation that one is less likely to find 
low$-$velocity pairs at larger separations, thereby implying 
that mergers should also be less probable.  

\section{DISCUSSION}\label{ssrs2mr:discuss}

\subsection{Comparison with Earlier Estimates of the 
Local Pair Fraction}
All published estimates of the local pair fraction have been 
hindered by small sample sizes and a lack of redshifts.  
In addition, as demonstrated throughout this paper, the 
traditional pair fraction is not a robust statistic, particularly 
when applied to flux-limited surveys.  The new statistics 
introduced in this paper, along with 
careful accounting for selection effects such as the 
flux limit, yield the first secure measures of pair statistics
at low redshift.  Therefore, strictly 
speaking, the results in this paper cannot be compared directly 
with earlier pair statistics.  However, 
it is possible to check for general consistency in results, and 
we will attempt to do so.  

As discussed in Section~\ref{ssrs2mr:background}, 
Patton et al. (1997) estimated the local pair 
fraction to be $4.3 \pm 0.4\% $, using the UGC catalog.  
The Patton et al. (1997) estimate was based on a flux-limited 
sample with 
$B \leq 14.5$, and a mean redshift of $z$=0.0076. 
This corresponds roughly to an average limiting absolute magnitude 
of $M_B$ = $-17.3$.  Loosely speaking, this is analogous to $M_2$.  
The pair definition used in their estimate was 
$r_p \leq 20$ \hkpc, with no $\Delta v$ criterion.  
$N_c$ may be interpreted 
as an approximation to the traditional pair fraction, 
provided the relative proportion of triples is small.  
We recompute the SSRS2 pair statistics, 
using $\Delta \vmax = 1000$ km/s in an attempt to match the 
results that would be found using no $\Delta v$ criterion (see 
Figure~\ref{ssrs2mr:figrl}).  
We find $N_c = 0.026 \pm 0.006$.  
This implies a local pair fraction of $2.6 \pm 0.6\%$.  
This value is somewhat smaller than the earlier result, 
with larger errors.  
We strongly emphasize that, while these results are broadly 
consistent, we would not expect excellent agreement, due to 
the improved techniques used in this study.
 
\subsection{The Merger Fraction at $z\sim 0$}\label{ssrs2mr:fmg} 
The quantities $N_c$ and $L_c$ are practical measures of the average numbers
and luminosities of companions with relatively high merging
probabilities.  However, the
ambiguity of redshift space is such that some of these companions can be
entirely safe from ever merging.  That is, $\Delta v= 500$ km/s can 
correspond either to 
two galaxies at a small physical separation with a large
infall velocity, or, to two galaxies at a separation of 5\hmpc, with no
relative peculiar velocity.  In order to transform from $N_C$ (and
$L_C$) to an estimate of the incidence of mergers, we must determine
what fraction of our close companions have true 3-dimensional physical
separations of $r < 20$~\hkpc.  This quantity, which we refer to as
$f_{\rm 3D}$, has been discussed previously by Yee and Ellingson
(1995\nocite{YE}).  This quantity needs to be evaluated based on the
small separation clustering and kinematics of the galaxy population,
$r_0$, $\gamma$, and $\sigma_{12}$, the parameters $\Delta v$ and
$\rpmax$, and the projected separation at which two galaxies
``optically overlap'' in the image. The quantity $f_{\rm 3D}$ is then
evaluated with a triple integral, first placing the correlation
function into redshift space, then integrating over projected and velocity
separation. For reasonable choices of pair selection parameters, the
outcome is fairly stable at $f_{\rm 3D}\simeq 0.5$. The most important
parameter is $\Delta V/\sigma_{12}$, which should lie in the range of 2 to
4.  If this parameter is too small, physical pairs will be missed; if
it is too large, too many distant companions will be incorporated.  
Other reasonable choices include setting the ratio of the overlap 
separation to maximum pair separation to be at least three, 
and the ratio of the maximum pair separation to correlation length 
to be at least a factor of 30.  We will take
$f_{\rm 3D}$=0.5 to be the best estimate currently available.

We can now estimate the merger fraction ($\fmg$) at the present 
epoch.  In this study, we have found $N_C$=0.0226$\pm 0.0052$.  
As most companions are found in pairs, rather than triplets 
or higher order N-tuples, this is comparable to the 
fraction of galaxies in close pairs.  From our estimate 
of $f_{\rm 3D}$, we infer that half of these galaxies are 
in merging systems, yielding $\fmg\approx 0.011$.  
This implies that approximately 1.1\% of $-21 \leq M_B \leq -18$ 
galaxies are undergoing mergers at the present epoch.  
We stress here that this result applies only to 
galaxies within the specified absolute magnitude limits.  
Probing to fainter luminosities would cause $\fmg$ to 
increase substantially.  In addition,  this result  
applies only to the close companions defined in 
this analysis.  Clearly, by modifying this definition (and 
therefore changing the typical merger timescale under 
consideration), the merger fraction would also be 
certain to change.  

\subsection{The Merger Timescale}\label{ssrs2mr:tmg}
We now have an idea of how prevalent ongoing mergers 
are at the present epoch.  
In order to relate this result to 
the overall importance of mergers, 
we must estimate the merger timescale ($\tmg$).  
We will use the properties of our SSRS2 pairs to estimate 
the mean dynamical friction timescale for pairs in our 
sample.  
Following Binney and Tremaine (1987\nocite{BT}), 
we assume circular orbits and a dark matter density profile 
given by $\rho(r)\propto r^{-2}$.  The dynamical friction 
timescale $T_{\rm fric}$ (in Gyr) is given by 
\begin{equation}\label{ssrs2mr:eqntfric}
T_{\rm fric}={2.64\times 10^5 r^2 v_c \over M \ln \Lambda},
\end{equation}
where $r$ is the initial physical pair separation in kpc, 
$v_c$ is the circular velocity in km/s, $M$ is the mass ($\msun$), 
and $\ln \Lambda$ is the Coulomb logarithm.
We estimate $r$ and $v_c$ using the pairs in 
Tables~\ref{ssrs2mr:tabcpn} and~\ref{ssrs2mr:tabcps}.  
The mean projected separation is $r_p\sim 14$\hkpc.  
As our procedure already includes a correction from projected 
separation ($r_p$) to 3-dimensional separation ($r$), 
we take $r=r_p$.  Assuming $h$=0.7, this leads to 
$\bar{r}\sim 20$ kpc.  
The mean line of sight 
velocity difference is $\Delta v \sim 150$ km/s.  
We assume the velocity distribution is isotropic, which 
implies that $v_c = \sqrt{3}\Delta v \sim 260$ km/s.  
The mean absolute magnitude of companions is $M_B\sim -19$ 
(see Figure~\ref{ssrs2mr:figlh}).  We assume a representative estimate of 
the galaxy mass-to-light ratio of $M/L\sim 5$, yielding 
a mean mass of $M \sim 3 \times 10^{10} \msun$.  
Finally, Dubinski, Mihos, and Hernquist (1999)\nocite{DMH99} 
estimate $\ln \Lambda \sim 2$, which fits the orbital decay 
of equal mass mergers seen in simulations.   
Using Equation~\ref{ssrs2mr:eqntfric}, we find 
$T_{\rm fric}\sim 0.5$ Gyr.  We caution that this estimate 
is an approximation, and is averaged over systems with 
a wide range in merger timescales.  Nevertheless, we 
will take $\tmg$ = 0.5 Gyr as being representative of 
the merger timescale for the pairs in our sample.  

\subsection{The Cumulative Effect of Mergers Since $z \sim 1$}
\label{ssrs2mr:frem}
Now that we have estimated the present epoch merger fraction 
and the merger timescale, we will attempt to ascertain 
what fraction of present galaxies have undergone mergers 
in the past.  These galaxies can be classified as 
merger remnants; hence, we will refer to this fraction 
as the remnant fraction ($\frem$).  
We begin by imagining the state of affairs at a lookback 
time of $t=\tmg$.  
Suppose the merger fraction at 
the corresponding redshift is given by $\fmg(z)$.  
In the time interval between then and the present, 
a fraction $\fmg(z)$ of galaxies will undergo mergers, 
yielding $0.5\fmg(z)$ merger remnants.  Therefore, the
remnant fraction at the present epoch is given by 
\begin{equation}\label{ssrs2mr:eqnrem}
\frem = 1 - {1- \fmg(z) \over 1 - 0.5 \fmg(z)}.
\end{equation}
Similarly, if we extend this to a lookback time of 
$N\tmg$, where $N$ is an integer, then the remnant 
fraction is given by 
\begin{equation}\label{ssrs2mr:eqnremz}
\frem = 1 - \prod_{j=1}^N{1- \fmg(z_j) \over 1 - 0.5 \fmg(z_j)},
\end{equation}
where $z_j$ corresponds to a lookback time of $t=j\tmg$.  
We now make the simple assumption that the merger rate 
does not change with time.  In this case, our present epoch 
estimate of the merger fraction holds at all redshifts, 
giving $\fmg(z)$=0.011.  In order to convert between 
redshift and lookback time, we must specify a cosmological 
model.  We assume a Hubble constant of $h$=0.7.  
For simplicity, we assume $q_0$=0.5.  
Therefore, $z=(1-3H_0t/2)^{-2/3}-1$.  Using our merger timescale 
estimate of $\tmg$=0.5 Gyr, we can now investigate the 
cumulative effect of mergers.  With the chosen cosmology, 
$z$=1 corresponds to a lookback time of $\sim$ 6 Gyr, 
or 12$\tmg$ ($N$=12).  With this lookback time, 
Equation~\ref{ssrs2mr:eqnremz} 
yields $\frem$=0.066.  
This implies that $\sim$ 6.6\% of galaxies 
with $-21 \leq M_B \leq -18$ have undergone mergers since 
$z\sim 1$.  

If the mergers taking place in our sample produce elliptical 
galaxies, it is worthwhile comparing the remnant fraction 
to the elliptical fraction (cf. Toomre 1977\nocite{T77}).  
The elliptical fraction for bright field galaxies is generally 
found to be about 10\% (e.g., Dressler 1980\nocite{D80}, 
Postman and Geller 1984\nocite{PG84}).  This result is 
broadly consistent with the remnant fraction found in 
this study.  

While our estimate of the remnant fraction 
is based on our statistically secure 
measurement of $N_C$, it also relies on fairly crude 
assumptions regarding the merger fraction and merger timescale.  
In particular, the merger rate has been assumed to be 
constant.  There is no physical basis for this assumption; 
in fact, a number of studies have predicted a rise in the 
merger rate with redshift.  If this is true, we will have 
underestimated the remnant fraction, and the relative 
importance of mergers.  
In a future paper (Patton et al. 2000\nocite{P00}), 
we will address this issue 
by investigating how the merger rate changes with redshift.  

\section{CONCLUSIONS}\label{ssrs2mr:conclude}
We have introduced two new pair statistics, $N_c$ and $L_c$, 
which are shown to be related to the galaxy merger and 
accretion rates respectively.  Using Monte Carlo simulations, 
these statistics are found to be robust to the redshift-dependent 
density changes 
inherent in flux-limited samples; this represents a very 
significant improvement over all previous estimators.  
In addition, we provide a clear prescription for relating 
$N_c$ and $L_c$ to the galaxy CF and 
LF, enabling straightforward comparison 
with measurements on larger scales.  

These statistics are applied to the SSRS2 survey, providing 
the first statistically sound measurements of pair statistics at 
low redshift.  For an effective range in absolute magnitude 
of $-21 \leq M_B \leq -18$, we find $N_c = 0.0226 \pm 0.0052$
at $z$=0.015, implying 
that $\sim$ 2.3\% of these galaxies have companions within 
a projected physical separation of 
5 \hkpc~ $\leq r_p \leq$ 20 \hkpc~ 
and 500 km/s along the line of sight.  
If this pair statistic remains fixed with redshift, 
simple assumptions imply that $\sim$ 6.6\% of present 
day galaxies with $-21 \leq M_B \leq -18$ have undergone 
mergers since $z$=1. 
For our luminosity statistic, we find 
$L_c = 0.0216 \pm 0.0055 \times 10^{10}~h^2 \lsun$.  
This statistic gives the mean luminosity in companions, per galaxy.
Both of these statistics will serve as local benchmarks in 
ongoing and future studies aimed at detecting 
redshift evolution in the galaxy merger and accretion rates.

It is our hope that these techniques will be applied to 
a wide range of future redshift surveys.  
As we have demonstrated, one must 
carefully account for differences in sampling effects 
between pairs and field galaxies.  This will be of 
increased importance when applying pair statistics 
at higher redshift, as $k$-corrections, luminosity 
evolution, band-shifting effects, 
and spectroscopic completeness have to be properly accounted 
for.  
The general approach outlined in this paper indicates the 
steps that must be taken to allow for a fair  
comparison between disparate surveys at low 
and high redshift.  
These techniques are applicable to 
redshift surveys with varying degrees of completeness, and 
are also adaptable to redshift surveys with additional 
photometric information, such as photometric redshifts, or 
even simply photometric identifications.  Finally, this 
approach can be used for detailed 
studies of both major and minor mergers.  

\acknowledgments
We wish to thank all members of the SSRS2 colloboration 
for their work in compiling the SSRS2 survey, and for 
making these data available in a timely manner.  
Digitized Sky Survey images used in this research were 
obtained from the Canadian Astronomical Data Centre (CADC), 
and are based on photographic data of the National 
Geographic Society Palomar Observatory Sky Survey (NGS-POSS).  
The Digitized Sky Surveys were produced at the Space Telescope 
Science Institute under U.S. Government grant NAG W-2166. 
This work was supported by the
Natural Sciences and Engineering Research Council of Canada, 
through research grants to R.G.C. and C.J.P..

\begin{table}[h]
\caption{SSRS2 Pair Statistics \label{ssrs2mr:tabstats}}
\begin{center}
\begin{tabular}[t]{|c|c|c|c|c|c|}
\hline
Sample&N&N$_{comp}$&$\bar{z}$&$N_c$&$L_c (10^{10} h^2 \lsun)$\\
\hline
SSRS2 North&  1702&  27& 0.014& 0.0189$\pm$ 0.0074& 0.0169$\pm$ 0.0079\\
SSRS2 South&  3067&  53& 0.016& 0.0261$\pm$ 0.0073& 0.0262$\pm$ 0.0078\\
SSRS2 (N+S)&  4769&  80& 0.015& 0.0226$\pm$ 0.0052& 0.0216$\pm$ 0.0055\\
\hline
\end{tabular}
\end{center}
\end{table}


\begin{table}[h]
\caption{SSRS2 North : Close Pairs and Triples \label{ssrs2mr:tabcpn}}
\begin{center}
\begin{tabular}[t]{|c|c|r|r|c|c|r|}
\hline
ID&N&$r_p~$&$\Delta v$&RA(1950.0)&DEC(1950.0)&$cz$~~~\\
\hline
1&2&17.3&75&11:18:00.6&-09:59:44&1523\\
2&2&12.7&385&12:23:39.7&-07:24:21&5183\\
3&2&19.8&392&12:39:42.8&-05:30:33&6604\\
4&2&18.1&357&12:49:52.3&-15:14:45&4539\\
5&2&13.0&183&12:50:31.4&-08:55:47&4149\\
6&3&11.4&42&13:01:21.2&-11:13:39&2965\\
7&2&12.3&227&13:17:00.1&-27:09:24&1835\\
8&2&6.5&198&13:23:14.6&-26:24:59&9596\\
9&2&18.1&99&13:30:31.8&-15:52:26&5813\\
10&2&12.5&70&13:32:37.2&-10:37:55&6550\\
11&2&15.7&298&13:57:36.5&-23:03:54&10592\\
12&2&12.8&4&14:01:52.3&-24:35:25&2117\\
13&2&17.4&90&14:10:41.5&-02:56:41&1664\\
\hline
\end{tabular}
\end{center}
\end{table}


\begin{table}[h]
\caption{SSRS2 South : Close Pairs and Triples \label{ssrs2mr:tabcps}}
\begin{center}
\begin{tabular}[t]{|c|c|r|r|c|c|r|}
\hline
ID&N&$r_p$~&$\Delta v$&RA(1950.0)&DEC(1950.0)&$cz$~~~\\
\hline
14&2&7.8&48&00:13:51.2&-06:35:57&8070\\
15&2&13.7&276&00:56:52.5&-05:04:07&5771\\
16&2&18.9&100&01:06:26.0&-28:51:25&5522\\
17&2&14.1&163&01:10:01.2&-04:24:30&5723\\
18&2&12.4&160&01:13:05.7&-06:51:23&6365\\
19&2&12.4&229&01:16:22.5&-17:04:00&6002\\
20&2&16.8&114&01:34:10.7&-37:35:09&5171\\
21&2&10.7&173&02:06:55.3&-10:22:13&4143\\
22&2&6.9&39&02:26:44.0&-11:03:15&4695\\
23&2&19.6&102&02:55:45.2&-10:33:21&4647\\
24&2&16.0&78&03:04:11.5&-39:13:15&6188\\
25&2&19.2&115&03:08:51.0&-09:06:01&4020\\
26&2&12.9&18&03:22:17.8&-03:12:17&2764\\
27&2&5.4&43&04:10:06.5&-32:59:21&1085\\
28&3&7.8&61&04:13:09.8&-28:36:32&5450\\
29&2&10.8&256&04:29:34.7&-29:51:29&4032\\
30&3&19.7&227&21:08:47.2&-23:22:03&10682\\
31&2&10.8&99&21:59:10.7&-32:13:26&2674\\
32&2&11.9&66&22:03:22.9&-28:11:55&7042\\
33&2&16.8&59&22:19:05.8&-25:50:09&6904\\
34&2&14.2&107&22:52:34.7&-34:09:30&8700\\
35&2&19.6&36&22:55:33.0&-04:02:30&3828\\
36&2&14.0&129&22:57:31.1&-13:05:22&3115\\
37&2&6.5&47&23:00:35.1&-09:15:38&7360\\
38&2&19.4&409&23:23:42.7&-39:29:48&10759\\
39&2&15.5&86&23:45:09.7&-28:24:56&8700\\
\hline
\end{tabular}
\end{center}
\end{table}


\begin{table}[h]
\caption{SSRS2 Pair Statistics for Various Choices of $M_2(B)$
\label{ssrs2mr:tabm2}}
\begin{center}
\begin{tabular}[t]{|c|c|c|}
\hline
$M_2$&$N_c$&$L_c (10^{10} h^2 \lsun)$\\
\hline
 -19.0& 0.0082$\pm$ 0.0019& 0.0135$\pm$ 0.0034\\
 -18.9& 0.0093$\pm$ 0.0021& 0.0144$\pm$ 0.0037\\
 -18.8& 0.0105$\pm$ 0.0024& 0.0154$\pm$ 0.0039\\
 -18.7& 0.0117$\pm$ 0.0027& 0.0163$\pm$ 0.0042\\
 -18.6& 0.0131$\pm$ 0.0030& 0.0171$\pm$ 0.0044\\
 -18.5& 0.0145$\pm$ 0.0033& 0.0180$\pm$ 0.0046\\
 -18.4& 0.0159$\pm$ 0.0037& 0.0188$\pm$ 0.0048\\
 -18.3& 0.0175$\pm$ 0.0040& 0.0195$\pm$ 0.0050\\
 -18.2& 0.0191$\pm$ 0.0044& 0.0203$\pm$ 0.0052\\
 -18.1& 0.0208$\pm$ 0.0048& 0.0210$\pm$ 0.0054\\
 -18.0& 0.0226$\pm$ 0.0052& 0.0216$\pm$ 0.0055\\
 -17.9& 0.0244$\pm$ 0.0056& 0.0222$\pm$ 0.0057\\
 -17.8& 0.0263$\pm$ 0.0060& 0.0228$\pm$ 0.0058\\
 -17.7& 0.0282$\pm$ 0.0065& 0.0234$\pm$ 0.0060\\
 -17.6& 0.0302$\pm$ 0.0070& 0.0239$\pm$ 0.0061\\
 -17.5& 0.0323$\pm$ 0.0074& 0.0244$\pm$ 0.0062\\
 -17.4& 0.0344$\pm$ 0.0079& 0.0249$\pm$ 0.0064\\
 -17.3& 0.0366$\pm$ 0.0084& 0.0253$\pm$ 0.0065\\
 -17.2& 0.0388$\pm$ 0.0089& 0.0257$\pm$ 0.0066\\
 -17.1& 0.0411$\pm$ 0.0095& 0.0261$\pm$ 0.0067\\
 -17.0& 0.0435$\pm$ 0.0100& 0.0264$\pm$ 0.0068\\
\hline
\end{tabular}
\end{center}
\end{table}

\newpage



\begin{figure}
\fbox{See f6a.gif, f6b.gif, f6c.gif}
\figcaption[figima.ps]
{
A mosaic of images is given for the 38 
close (5 \hkpc~$ < r_p \leq 20$ \hkpc) dynamical 
($\Delta V < 500$ km/s) pairs or triples satisfying the criteria  
used for computing pair statistics.  
These images were obtained from the Digitized Sky Survey.  
Each image is 50~\hkpc~ on a side, corresponding to angular sizes 
of 1.5\arcmin$-$10\arcmin.  
\label{ssrs2mr:figim}
}
\end{figure}



\end{document}